\documentstyle[aps,preprint,tighten,floats,epsfig]{revtex}
\begin{document}
\newcommand{\nc}{\newcommand}

\nc{\bib}{\bibitem}
\nc{\1}{16^3\times{32}}
\nc{\2}{24^3\times{36}}
\nc{\al}{\alpha}
\nc{\napefv}{n_{\rm{ape}}(0.55)}
\nc{\nausfv}{n_{\rm{aus}}(0.55)}
\nc{\napeprm}{n_{\rm{ape}}(\alpha')}
\nc{\nausprm}{n_{\rm{aus}}(\alpha')}
\nc{\ncl}{n_c}
\nc{\nape}{n_{\rm{ape}}(\alpha)}
\nc{\naus}{n_{\rm{aus}}(\alpha)}
\nc{\napep}{n_{\rm{ape}}(\alpha')}
\nc{\nausp}{n_{\rm{aus}}(\alpha')}
\nc{\sutwo}{SU_{c}(2)}
\nc{\suthree}{SU_{c}(3)}
\nc{\sun}{SU_{c}(N)}
\nc{\szero}{S_{0}}
\nc{\szr}{\szero=8\pi^2/g^2}
\nc{\ql}{Q_{\rm{L}}(x)}

\draft
\preprint{\vbox{\noindent\null \hfill ADP-00-03/T391\\ 
                         \null \hfill hep-lat/0001018 \\
}}

\title{\bf Calibration of Smearing and Cooling Algorithms in SU(3)-Color
Gauge Theory} 

\author{Fr\'{e}d\'{e}ric D.R. Bonnet\footnote{
E-mail:~fbonnet@physics.adelaide.edu.au ~$\bullet$~ Tel:
+61~8~8303--3428 ~$\bullet$~ Fax: +61~8~8303--3551}$^{1}$,
Patrick Fitzhenry\footnote{E-mail:~pfitzhen@physics.adelaide.edu.au}$^{1}$, 
Derek B. Leinweber\footnote{E-mail:~dleinweb@physics.adelaide.edu.au
~$\bullet$~ Tel: +61~8~8303--3423 ~$\bullet$~ Fax: +61~8~8303--3551
\hfill\break 
\null\quad\quad 
WWW:~http://www.physics.adelaide.edu.au/theory/staff/leinweber/}$^{1}$,
Mark R. Stanford\footnote{E-mail:~mstanfor@physics.adelaide.edu.au}$^{1}$,
Anthony G. Williams\footnote{E-mail:~awilliam@physics.adelaide.edu.au
~$\bullet$~ Tel: +61~8~8303--3546 ~$\bullet$~ Fax: +61~8~8303--3551
\hfill\break 
\null\quad\quad
WWW:~http://www.physics.adelaide.edu.au/cssm/}$^{1,2}$.}
\address{ $^1$ Special Research Center for the Subatomic Structure of
Matter (CSSM) and Department of Physics and Mathematical Physics,
University of Adelaide 5005, Australia.} 
\address{ $^2$ Department of Physics and SCRI, Florida State
University Tallahassee, FL 32306.} 
\date{\today}
\maketitle

\begin{abstract}
The action and topological charge are used to determine the relative
rates of standard cooling and smearing algorithms in pure SU(3)-color
gauge theory.  We consider representative gauge field configurations
on $16^3\times 32$ lattices at $\beta=5.70$ and $24^3\times 36$
lattices at $\beta=6.00$.  We find the relative rate of variation in
the action and topological charge under various algorithms may be
succinctly described in terms of simple formulae.  The results are in
accord with recent suggestions from fat-link perturbation theory.
\end{abstract}

\newpage

\section{Introduction}

Cooling \cite{cooling} is a well established method for locally
suppressing quantum fluctuations in gauge field configurations.  The
locality of the method allows topologically nontrivial field
configurations to survive numerous iterations of the cooling
algorithm.  Application of such algorithms are central to removing
short-distance fluctuations responsible for causing renormalization
constants to significantly deviate from one as is the
case for the topological charge operator \cite{campostrini}.

An alternative yet somewhat related approach is the application of APE
smearing \cite{ape} to the gauge field configurations.  Because the
algorithm is applied simultaneously to all link variables, without any
annealing in the process of a lattice update, the application of the
algorithm may be viewed as an introduction of higher dimension
irrelevant operators designed to improve the operator based on
unsmeared links.  This approach also leads to the desirable effect of
renormalization constants approaching one.  The net effect may be
viewed as introducing a form factor to gluon vertices
\cite{Bernard:1999kc} that suppresses large $q^2$ interactions and, as
a result, suppresses lattice artifacts at the scale of the cutoff.

The focus of this paper is to calibrate the relative rates of cooling
and smearing as measured by the action and the topological charge.  In
doing so we aim to establish relationships between various algorithms
and their internal parameters.  It is well known that errors in the
standard Wilson action eventually destroy (anti)instanton
configurations and this effect is represented in the topological
charge as sharp transitions from one integer to another.  We find the
timing analysis of these topological charge defects to be central to
understanding the behavior of the various algorithms.

To bridge the gap between cooling and smearing and discover the origin
of variations between these algorithms, we present a new algorithm for
smearing the gauge fields which we refer to as Annealed $U$ Smearing
or AUS smearing.  In addition to comparing cooling and smearing, we
also calibrate the effects of different smearing fractions, describing
the balance between the unsmeared link and the smeared link.

There have been numerous studies of the dependence of vacuum matrix
elements on the number of smoothing sweeps.  Of particular note is the
observation of cooling as a diffusive procedure \cite{giacomo97}, such
that the distance $d$ affected by cooling is related to the
square-root of the number of sweeps over the lattice, $n$.  This
random walk ansatz was further investigated by introducing the lattice
spacing $a$, such that $d = a \sqrt{n}$ takes on a physical length
\cite{ringwald99}.  With this simple formula, it is possible to
effectively relate results from lattices of different $\beta$ ($a$)
and different smoothing sweeps $n$.  Extensive work on removing the
dependence of particular quantities on the number of smoothing sweeps
was carried out in Ref.\ \cite{deforcrand}.  There, highly improved
actions are fine tuned to stabilize the instanton action and size
under cooling.  Moreover, a modification of the updating algorithm
\cite{garcia99} allows one to define a physical length scale beyond
which the physics is largely unaltered.  Closer to the focus of this
investigation, some exploratory comments favoring the possibility of
calibrating under-relaxed cooling were made in Ref.\ \cite{smith98}.
We will lend further weight to this possibility in the following
analysis.

This paper is divided as follows: Section~\ref{action}, briefly
describes the gauge action and summarizes the numerical simulations
examined in this study.  Section~\ref{algos} presents the relaxation
algorithms and topological charge definition on the lattice.
Sections~\ref{numact} and ~\ref{numtopo} present the numerical results
for the action and topological charge trajectories respectively.  A
summary of our findings is provided in Section~\ref{summary}.

\section{Lattice Gauge Action for $\suthree$}
\label{action}
\subsection{Gauge Action.}

The Wilson action is defined as,
\begin{equation}
S_G=\beta\sum_x \sum_{\mu < \nu}\frac{1}{3}{\cal R}e \, {\rm
Tr}(1-U_{\mu \nu}(x)), 
\label{gaugeaction}
\end{equation}
where the operator $U_{\mu\nu}(x)$ is the standard plaquette,
\begin{equation}
U_{\mu\nu}(x)=U_{\mu}(x)\, U_{\nu}(x+\hat{\mu})\,
U^{\dagger}_{\mu}(x+\hat{\nu})\, U^\dagger_{\nu}(x) \, .
\label{staples}
\end{equation}

Gauge configurations are generated using the
Cabbibo-Marinari~\cite{Cab82} pseudo--heat--bath algorithm with three
diagonal $\sutwo$ subgroups. Simulations are performed using a
parallel algorithm on a Thinking Machines Corporations (TMC) CM-5 with
appropriate link partitioning.  For the Wilson action we partition the
link variables in the standard checkerboard fashion.

Configurations are generated on a $16^3\times{32}$ lattice at
$\beta=5.70$ and a $24^3\times{36}$ lattice at $\beta=6.00$.
Configurations are selected after 5000 thermalization sweeps from a
cold start, and every 1000 sweeps thereafter.  Lattice parameters are
summarized in Table~\ref{simultab}.  When required, we use the
plaquette definition of the mean link variable.

\begin{table}[b]
\caption{Parameters of the numerical simulations.}
\begin{tabular}{cccccccc}
Action &Volume &$N_{\rm{Therm}}$ & $N_{\rm{Samp}}$ &$\beta$ &$a$ (fm) & $u_{0}$ & Physical Volume (fm)\\
\hline
Wilson & $16^3\times{32}$ & 5000 & 1000 & 5.70 & 0.18  & 0.86085 & $2.88^3\times{5.76}$ \\
Wilson & $24^3\times{36}$ & 5000 & 1000 & 6.00 & 0.10  & 0.87777 & $2.4^3\times{3.6}$ \\
\hline
\end{tabular}
\label{simultab}
\end{table}

\section{Cooling and Smearing Algorithms}
\label{algos}

Relaxation techniques have been the subject of a large amount of study
in lattice QCD over the past few years \cite{recentConfProc}.
Numerous algorithms have been employed to suppress the high-frequency
components of the gauge field configurations in investigating their
semi-classical content.  In the following three subsections we
consider three different algorithms that perform this task.

Perhaps it should be noted that cooling algorithms have been designed
to render instantons stable over many hundreds of sweeps.  For
example, De Forcrand {\it et al.}~\cite{deforcrand} have established a
five-loop improved gauge action that achieves this.  However, our
interest here is in using the defects in the topological charge evolution
under standard cooling and smearing algorithms based on the Wilson
gauge action to calibrate these algorithms.

\subsection{Cooling}

Standard cooling minimizes the action locally at each link update.
The preferred algorithm is based on the Cabbibo-Marinari~\cite{Cab82}
pseudo-heat-bath algorithm for constructing $\suthree$-color gauge
configurations.  A brief summary of a link update follows.  

At the $\sutwo$ level the algorithm is transparent.  An element of
$\sutwo$ may be parameterized as, $U = a_{0}I +
i\,\vec{a}\cdot\vec{\sigma}$, where $a$ is real and $a^{2}=1$.
Let $\widetilde{U}_{\mu}$ be one of the six staples associated with
creating the plaquette associated with a link $U_\mu$.
\begin{equation}
\widetilde{U}_{\mu} = U_{\nu}(x+\hat{\mu})\,
U^{\dagger}_{\mu}(x+\hat{\nu})\, U^\dagger_{\nu}(x) \, .
\label{tildeU}
\end{equation}
We define
\begin{eqnarray}
\sum_{\alpha=1}^6 \widetilde{U}_{\alpha} = k \overline{U}
&\hspace*{0.55cm}{\rm where}\hspace*{0.55cm}& \overline{U}\in
\sutwo\hspace*{0.55cm}{\rm and}\hspace*{0.55cm}
k^2\equiv{\det\left(\sum_{\alpha=1}^6\widetilde{U}_{\alpha}\right)} \, . 
\label{coolingk}
\end{eqnarray}
The feature of the sum of ${\sutwo}$ elements being proportional to
an ${\sutwo}$ element is central to the algorithm.  The local
${\sutwo}$ action is proportional to
\begin{equation}
{\cal R}e\, {\rm Tr}(1- U\overline{U}) \, ,
\end{equation}
and is locally minimized when ${\cal R}e\, {\rm Tr}(U\overline{U})$ is
maximized, i.e.\ when
\begin{equation}
{\cal R}e\, {\rm Tr}(U\overline{U})={\cal R}e\, {\rm Tr}(I) \, .
\end{equation}
which requires the link to be updated as
\begin{equation}
U {\longrightarrow} U^\prime = \overline{U}^{-1} =
\overline{U}^\dagger = \frac{\left(\sum^6_{\alpha=1}
\widetilde{U_\alpha} \right)^\dagger}{k} \, .
\label{uprime}
\end{equation}
At the ${\suthree}$ level, one successively applies this algorithm to
various ${\sutwo}$ subgroups of ${\suthree}$, with ${\sutwo}$
subgroups selected to cover the ${\suthree}$ gauge group.  

We explored cooling gauge field configurations using two diagonal
$\sutwo$ subgroups.  The cooling rate is slow and the action density
is not smooth, even after 50 sweeps over the lattice.  Addition of the
third diagonal $\sutwo$ subgroup provides acceptably fast cooling and
a smooth action density.  These $\sutwo$ subgroups are acting as a
covering group for the $\suthree$ gauge group. One would therefore
expect that repeated updates of $\sutwo$ subgroups will better
approximate the ultimate ${\suthree}$ link.

\subsection{Smearing}

Here we consider two algorithms for smearing the gauge links.  APE
smearing \cite{ape} is now well established.  Our alternate algorithm
AUS smearing is designed to remove instabilities of the APE algorithm
and provide an intermediate algorithm sharing features of both
smearing and cooling.

\subsubsection{APE Smearing}

APE smearing \cite{ape} is a gauge equivariant \cite{Hetrick:1998}
prescription for averaging a link $U_\mu(x)$ with its nearest
neighbors $U_\mu(x+\hat\nu),\ \nu \ne \mu$.  The linear combination
takes the form:
\begin{equation}
U_{\mu}(x)\longrightarrow{U_{\mu}^\prime(x)} = 
(1-\alpha)U_{\mu}(x)+\frac{\alpha}{6}\widetilde{\Sigma}^\dagger(x;\mu),
\label{apecomb}
\end{equation}
where $\widetilde{\Sigma}(x;\mu) = \left( \sum^6_{\nu=1}
\widetilde{U_\nu}(x) \right)_\mu$, is the sum of the six staples
defined in Eq.~(\ref{tildeU}).  The parameter $\alpha$ represents the
smearing fraction. The algorithm is constructed as follows: (i)
calculate the staples, $\widetilde{\Sigma}(x;\mu)$; (ii) calculate the
new link variable $U_{\mu}^\prime(x)$ given by
Eq.~(\ref{apecomb}) and then reunitarize $U_{\mu}^\prime(x)$; (iii)
once we have performed these steps for every link, the smeared
$U^\prime_{\mu}(x)$ are mapped into the original $U_{\mu}(x)$.  This
defines a single APE smearing sweep which can then be repeated.

The celebrated feature of APE smearing is that it can be realized as
higher-dimension operators that might appear in a fermion action for
example.  Indeed, ``fat link'' actions based on the APE algorithm 
are excellent candidates for an efficient action with improved
chiral properties \cite{DeGrand:1999gp}.  The parameter space of fat
link actions is described by the number of smearing sweeps $n_{\rm ape}$
and the smearing coefficient $\alpha$.  It is the purpose of this
investigation to explore the possible reduction of the dimension of
this parameter space from two to one.

\subsubsection{AUS Smearing}
\label{aussmear}

The Annealed $U$ Smearing (AUS) algorithm is similar to APE smearing
in that we take a linear combination of the original link and the
associated staples, $\widetilde{\Sigma}^\dagger(x;\mu)$.  However in
AUS smearing we take what was a single APE smearing sweep over all the
links on the lattice and divide it up into four partial sweeps based
on the direction of the links.  One partial sweep corresponds to an
update of all links oriented in one of the four Cartesian directions
denoted $\mu=1,..,4$.  Hence in a partial sweep we calculate the
reuniterized $U_{\mu}^\prime(x)$ for all $\mu$--oriented links and
update all of these links
$[U_{\mu}^\prime(x)\longrightarrow{U_{\mu}(x)}]$ at the end of each
partial sweep.

Thus the difference between AUS smearing and APE smearing is that in
APE smearing no links are updated until all four partial sweeps are
completed, while in AUS smearing, updated smeared information is
cycled into the calculation of the next link direction, a process
commonly referred to as annealing.  In a sense AUS smearing is between
cooling and APE smearing in that cooling updates one link at a time,
AUS smearing updates one Cartesian direction at a time, and APE
smearing updates the whole lattice at the same time.

As for APE smearing, the reunitarization of the $\suthree$ matrix is
done using the standard row by row orthonormalization procedure: begin
by normalizing the first row; then update the second row by
$row2=row2-(row2\cdot{row1})row1$; normalize the second row; finally
set $row3$ equal to the cross product of $row1$ and $row2$.

For $\sutwo$ gauge theory, cooling and smearing are identical for the
case of $\alpha = 1$ when the links are updated one at a time.  The
AUS algorithm changes the level of annealing and provides the
opportunity to alter the degree of smoothing.  At the same time the
AUS algorithm preserves the gauge equivariance of the APE algorithm
for $\suthree$ gauge theory \cite{Hetrick:1998}.  As we shall see,
it also removes the instability of the APE algorithm at large smearing
fractions $\alpha$.  This latter feature may provide a more efficient
method for finding the fundamental modular region of Landau Gauge in
trying to understand the effects of Gribov copies in gauge dependent
quantities such as the gluon-propagator \cite{GluonPropPapers}.

\subsection{Topological Charge}

We construct the lattice topological charge density operator analogous
to the standard Wilson action via the clover definition of
${F}_{\mu\nu}$.  The topological charge $Q_{\rm{L}}$ and the
topological charge density $q_{\rm{L}}(x)$ are defined as follows
\begin{equation}
Q_{\rm{L}} {\equiv} \sum_{x}{q_{\rm{L}}(x)} = \frac{g^2}{32\pi^2}\,
\epsilon_{\mu\nu\rho\sigma}\, \sum_{x} \, {\rm Tr}({F}_{\mu\nu}(x)\,
{F}_{\rho\sigma}(x)) \, . 
\label{latticecharge}
\end{equation}
where the field strength tensor is 
\begin{eqnarray}
a^2 \, g \, F_{\mu\nu}(x) = -\frac{i}{8} \left[ \left( {\cal
O}_{\mu\nu}(x) - {\cal O}_{\mu\nu}^{\dagger}(x) \right) - \frac{1}{3}
\, {\rm Tr}\left( {\cal O}_{\mu\nu}(x) - {\cal O}_{\mu\nu}^{\dagger}(x) \right)
\right] + {\cal O}(a^4) \, ,
\end{eqnarray}
and ${\cal O}_{\mu\nu}(x)$ is
\begin{eqnarray}
{\cal O}_{\mu\nu}(x) &=& U_{\mu}(x)\, U_{\nu}(x+\hat{\mu})\,
	     U^{\dagger}_{\mu}(x+\hat{\nu}) \, U^\dagger_{\nu}(x) 
             \nonumber \\
	     &+& U_{\nu}(x)\,
	     U^{\dagger}_{\mu}(x+\hat{\nu}-\hat{\mu})\, 
             U^{\dagger}_{\nu}(x-\hat{\mu})\, U_{\mu}(x-\hat{\mu})
	     \nonumber \\ 
	     &+& U_{\mu}(x-\hat{\mu})\,
	     U^{\dagger}_{\nu}(x-\hat{\mu}-\hat{\nu})\,
	     U_{\mu}(x-\hat{\mu}-\hat{\nu})\, U_{\nu}(x-\hat{\nu})
	     \nonumber \\ 
	     &+& U^{\dagger}_{\nu}(x-\hat{\nu})\,
	     U_{\mu}(x-\hat{\nu})\, U_{\nu}(x+\hat{\mu}-\hat{\nu})\,
	     U^{\dagger}_{\mu}(x) \, .
\label{Omunu}
\end{eqnarray}
Lattice operators possess a multiplicative lattice renormalization
factor, $Q_L = {\cal Z}_{Q}(\beta) \, Q$ which relates the lattice
quantity $Q_L$ to the continuum quantity $Q$.  Perturbative
calculations indicate ${\cal Z}_{Q}(\beta) \approx 1 - 5.451/\beta +
{\cal O}(1/{\beta^2})$ \cite{campostrini}.  This large renormalization
causes a problem when one is working at $\beta\approx6.0$, as ${\cal
Z}_Q(\beta)\ll{1}$.  This implies that the topological charge is
almost impossible to calculate directly.  However when one applies
cooling or smearing techniques to remove the problem of the short
range quantum fluctuations giving rise to ${\cal Z}_Q(\beta)\ll{1}$,
one can resolve near integer topological charge.  One can apply the
operator $Q_L$ to cooled configurations or smeared configurations.
The latter case may be regarded as employing an improved operator in
which the smeared links are understood to give rise to additional
higher dimension irrelevant operators designed to provide a smooth
approach to the continuum limit.

\section{Numerical Simulations}

We analyzed two sets of gauge field configurations generated using the
Cabbibo-Marinari~\cite{Cab82} pseudo-heat-bath algorithm with three
diagonal $\sutwo$ subgroups as described in Table~\ref{simultab}.
Analysis of a few configurations proves to be sufficient to resolve
the nature of the algorithms in question. The two sets are composed of
ten $16^3\times 32$ configurations and five $24^3\times 36$
configurations.  For each configuration we separately performed 200
sweeps of cooling, 200 sweeps of APE smearing at four values of the
smearing fraction and 200 sweeps of AUS smearing at six values of the
smearing fraction.  

For clarity, we define the number of times an algorithm is applied to
the entire lattice as $\ncl$, ${\nape}$ and $\naus$ for cooling, APE
smearing and AUS smearing respectively.  We monitor both the total
action normalized to the single instanton action $S_0 = 8 \pi^2 / g^2$
and the topological charge $\ql$ of Eq.~(\ref{latticecharge}) and
observe their evolution as a function of the appropriate sweep
variable and smearing fraction $\alpha$.

For APE smearing we consider four different values for the smearing
fraction $\alpha$ including 0.30, 0.45, 0.55, and 0.70.  Larger
smearing fractions reveal an instability in the APE algorithm where
the links are rendered to noise.  

The origin of this instability is easily understood in fat-link
perturbation theory \cite{Bernard:1999kc} where the smeared vector
potential after $n$ APE smearing steps is given by
\begin{equation}
A_\mu^{(n+1)} (q) = \sum_\nu \left [ f^n(q) \,
\left ( \delta_{\mu \nu} - {\widehat q_\mu \widehat q_\nu \over
\widehat q^2} \right ) +
{\widehat q_\mu \widehat q_\nu \over \widehat q^2} \right ] 
A_\nu^{(n)} (q) \, ,
\end{equation}
reflecting the transverse nature of APE smearing.  Here
\begin{equation}
\widehat q_\mu = {2 \over a} \sin \left ( {a \, q_\mu \over 2} \right
)
\end{equation}
and
\begin{equation}
f(q) = 1 - {\alpha \over 6} \, \widehat q^2 \, .
\end{equation}
For $f(q)$ to act as a form factor at each vertex in perturbation
theory over the entire Brillouin zone
\begin{equation}
-{\pi \over a} < q_\mu \le {\pi \over a} \, ,
\end{equation}
one requires $-1 \le f(q) \le 1$ which constrains $\alpha$ to the
range $0 \le \alpha \le 3/4$.

The annealing process in AUS smearing removes this instability.
Hence, the parameter set for AUS smearing consists of the APE set plus
an extra two, $\al=0.85$ and 1.00.

\subsection{The Action Analysis.}
\label{numact}

\subsubsection{APE and AUS Smearing Calibration}

\begin{figure}[tp]
\centering{\
\epsfig{angle=90,figure=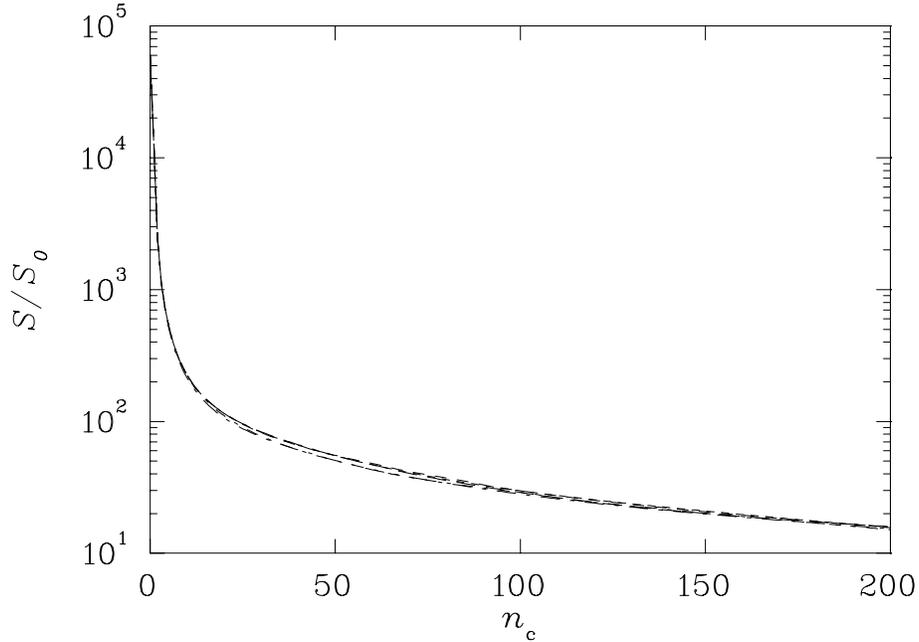,width=12cm}}
\caption{The ratio $S/S_0$ as a function of cooling sweeps $\ncl$ for
five configurations on the $\2$ lattice at $\beta=6.00$. The single
instanton action is $\szr$.
\label{accoolevolve}}
\end{figure}

The action normalized to the single instanton action $S/S_0$ can
provide some insight into the number of instantons left in the lattice
as a function of the sweep variable for each algorithm.  However, our
main concern is the relative rate at which the algorithms perform.  In
Fig.~\ref{accoolevolve}, we show $S/S_0$ as a function of cooling
sweep on the $\2$ lattice for five different configurations.  The
close proximity of the five curves is typical of the configuration
dependence of the normalized action $S/S_0$.  

\begin{figure}[tp]
\centering{\
\epsfig{angle=90,figure=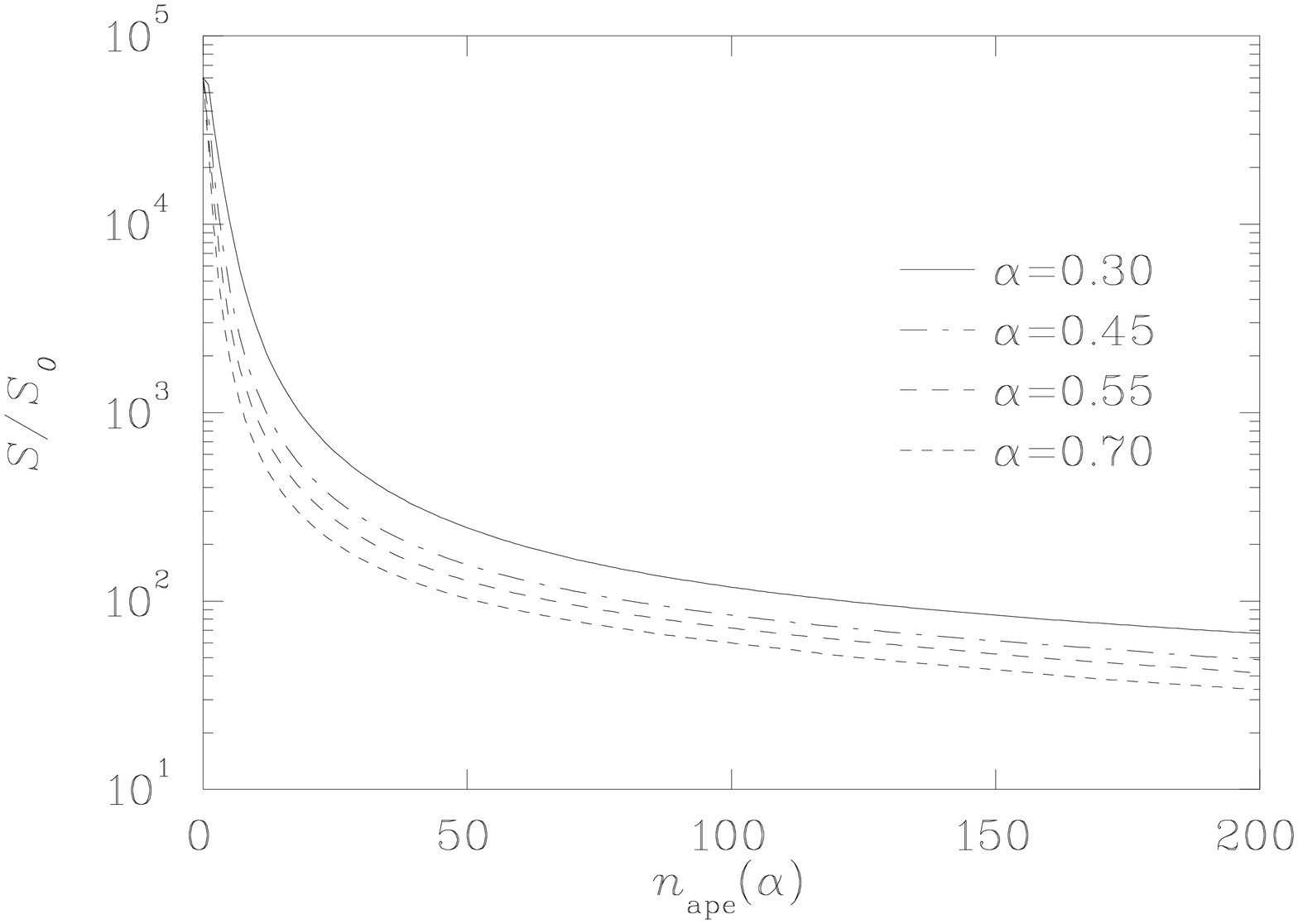,width=12cm}}
\caption{The ratio $S/\szero$ as a function of APE smearing sweeps
$\nape$ for a configuration on a $24^3\times{36}$ lattice at
$\beta=6.00$.  Each curve has an associated smearing fraction
$\alpha$. 
\label{acapeevolve}}
\end{figure}

\begin{figure}[tp]
\centering{\
\epsfig{angle=90,figure=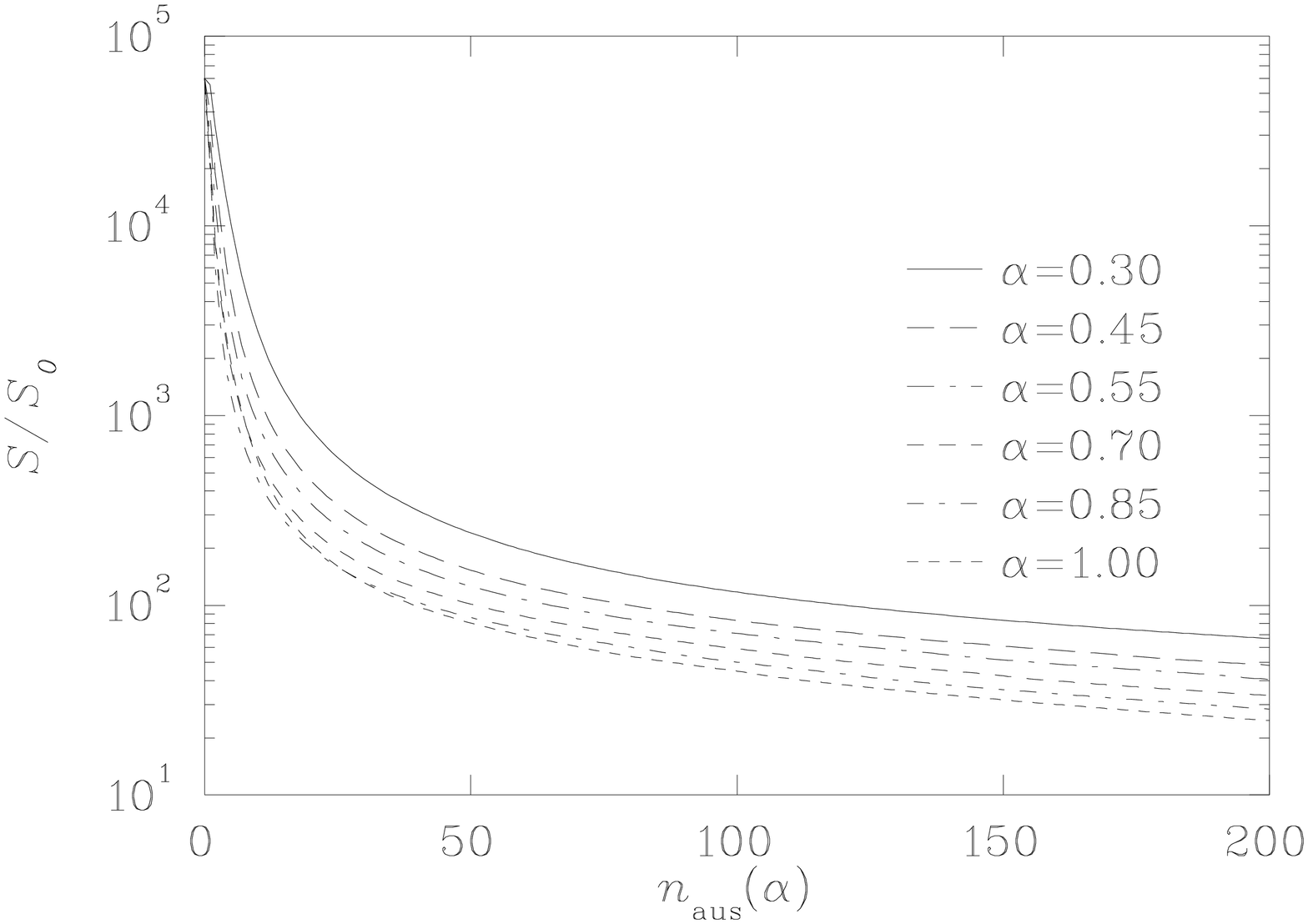,width=12cm}}
\caption{The ratio $S/\szero$ as a function of AUS smearing sweep
$\naus$ for a configuration on a $24^3\times{36}$ lattice at
$\beta=6.00$.  Each curve has an associated smearing fraction
$\alpha$.
\label{acausevolve}}
\end{figure}

Figures \ref{acapeevolve} and \ref{acausevolve} report results for
APE and AUS smearing respectively.  Here we focus on one of the five
configurations, noting that similar results are found for the other
configurations.  Each curve corresponds to a different value of the
smearing fraction $\alpha$.

A similar analysis of the $\1$ lattice at $\beta=5.70$ is also performed
yielding analogous results.  In fact, taking the physical volumes
of the two lattices into account reveals qualitatively similar action
densities after 200 sweeps. 

Note that in Fig.~\ref{acausevolve}, the curve associated with the
smearing parameter $\al=1.00$, crosses over the one generated at
$\al=0.85$, when the sweep number, $\naus$, is approximately 40
sweeps.  Thus $\alpha=1$ is not the most efficient smearing fraction
for untouched gauge configurations.  However, as the configurations
become smooth, $\al=1.00$ becomes the most efficient.

To calibrate the rate at which the algorithms reduce the action, we
record the number of sweeps $n(\alpha)$ required for the smeared
action to cross various thresholds $S_T$.  This is repeated for each
of the smearing fractions $\alpha$ under consideration.  In
establishing the relative $\alpha$ dependence for the number of sweeps
$n(\alpha)$, we first consider a simple linear relation between the
number of sweeps required to cross $S_T$ at one $\alpha$ compared to
another $\alpha'$, i.e.
\begin{equation}
n(\alpha') = c_0 + c_1 \, n(\alpha) \, .
\label{ansatz}
\end{equation}
Anticipating that $c_0$ will be small if not zero, we divide
both sides of this equation by $n(\alpha)$ and plot as a function of
$n(\alpha)$.  Deviations from a horizontal line will indicate failings
of our linear assumption.

\begin{figure}[tp]
\centering{\
\epsfig{angle=90,figure=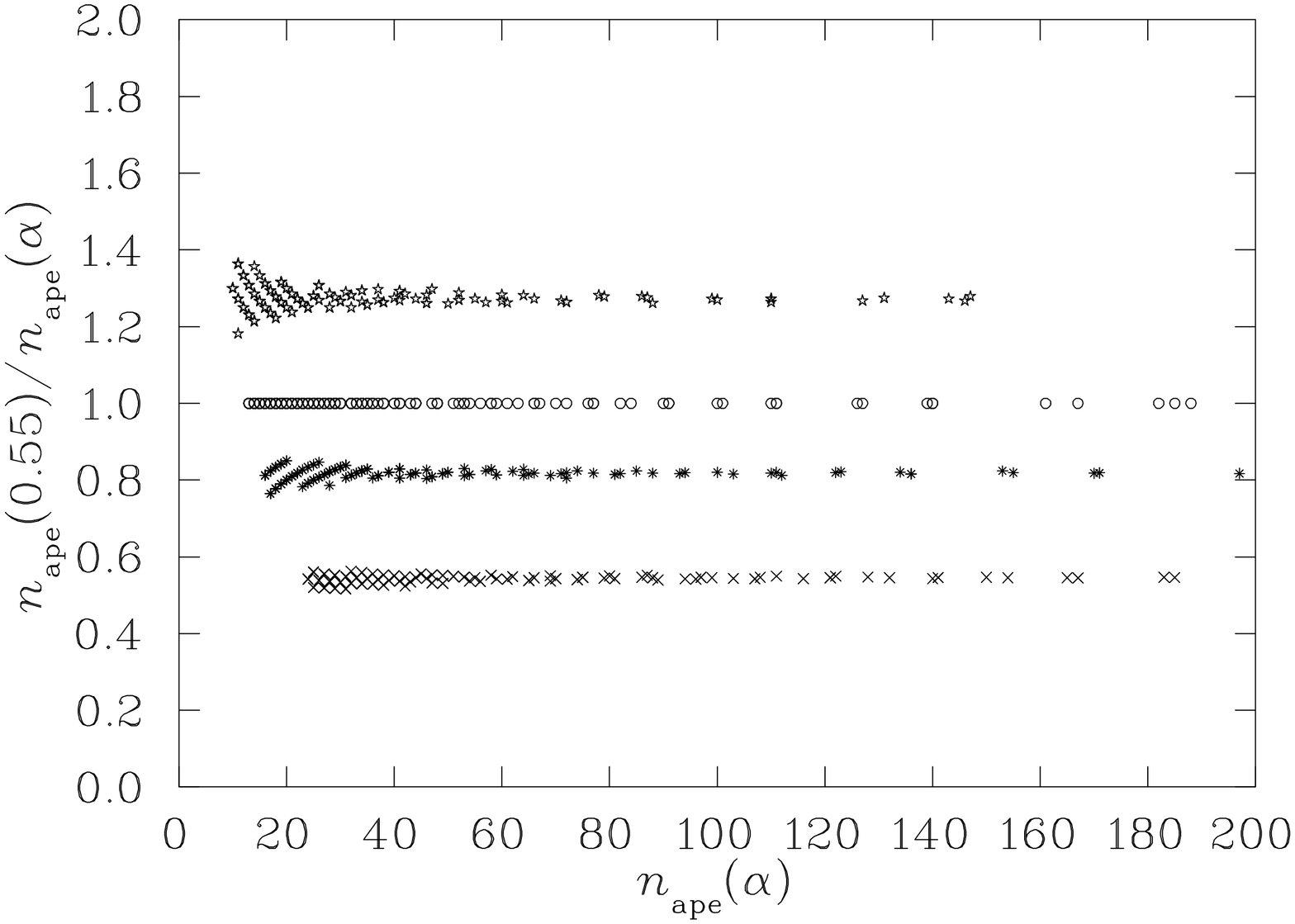,width=12cm}}
\caption{The ratio $\napefv/\nape$ versus $\nape$ for numerous
threshold actions on the $\2$ lattice at $\beta=6.00$.  From top down
the data points correspond to $\al = 0.70$, 0.55, 0.45 and 0.30.
\label{actape055vsal}}
\end{figure}

\begin{figure}[tp]
\centering{\
\epsfig{angle=90,figure=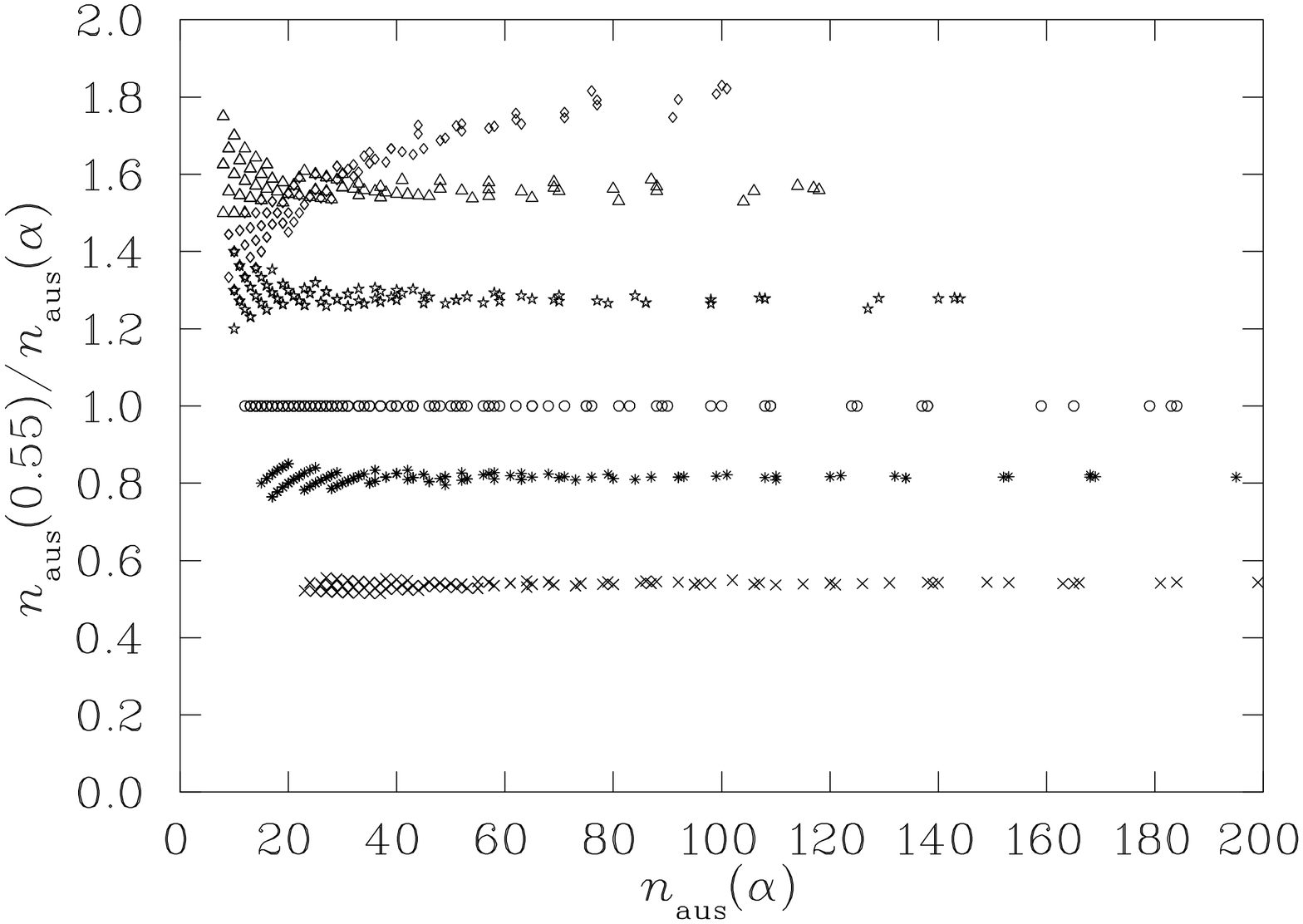,width=12cm}}
\caption{The ratio $\nausfv/\naus$ versus $\naus$ for numerous
threshold actions on the $\2$ lattice at $\beta=6.00$.  From top down
the data points correspond to $\al = 1.00$, 0.85, 0.70, 0.55, 0.45,
0.30.
\label{actaus055vsal}}
\end{figure}

Fig.~\ref{actape055vsal} displays results for $\alpha'$ fixed to
0.55 for the APE smearing algorithm and Fig.~\ref{actaus055vsal}
displays analogous results for AUS smearing.  We omit thresholds that
result in $n(0.55) < 10$ as these points will have integer
discretization errors exceeding 10\%.  For $\al = 0.70$ discretization
errors the order of 10\% are clearly visible in both figures.

For the smearing fraction $\alpha \le 0.85$ both plots show little
dependence of $n(\alpha'=0.55)/n(\alpha)$ on $n(\alpha)$.  This
supports our simple ansatz of Eq.~(\ref{ansatz}) and indicates $c_0$ is
indeed small as one would expect. 

The non-linear behavior for $\al=1.00$ in AUS smearing, in
Fig.~\ref{actaus055vsal}, reflects the cross over in
Fig.~\ref{acausevolve}, for $\al=1.00$.  For unsmeared
configurations, $\al=1.00$ appears to be too large.  Further study in
$\sutwo$ is required to resolve whether the origin of the smearing
inefficiency lies in the sum of staples lying too far outside the
$\suthree$ gauge group for useful reunitarization, or whether the
annealing of the links in AUS smearing is insufficient relative to
cooling.

To determine the $\alpha$ dependence of $c_1$, we plot $c_1 = \langle
n(\alpha'=0.55)/n(\alpha) \rangle$ as a function of $\alpha$.  Here
the angular brackets denote averaging over all the threshold values
considered; i.e. averaging over data in the horizontal lines of
Figs.~\ref{actape055vsal} and \ref{actaus055vsal}.

\begin{figure}[tp]
\centering{\
\epsfig{angle=90,figure=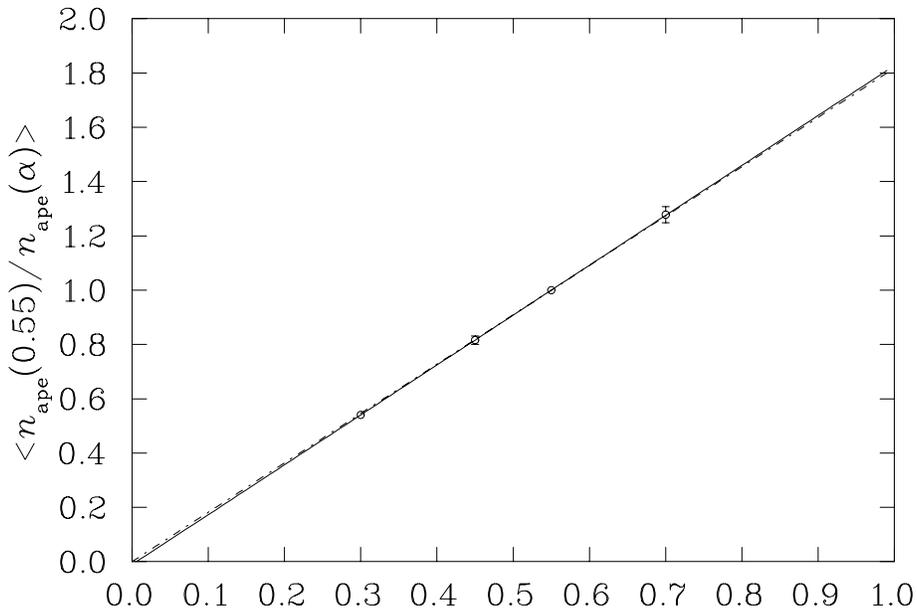,width=12cm}}
\caption{Illustration of the $\alpha$ dependence of $c_1 = \langle
n(\alpha'=0.55)/n(\alpha) \rangle$ for APE smearing.  The solid line
fit to the data indicates $c_1 = 1.838\, \alpha - 0.011$ whereas the
dashed line, constrained to pass through the origin, provides a slope
of 1.818.
\label{actaveapefit}}
\end{figure}

\begin{figure}[tp]
\centering{\
\epsfig{angle=90,figure=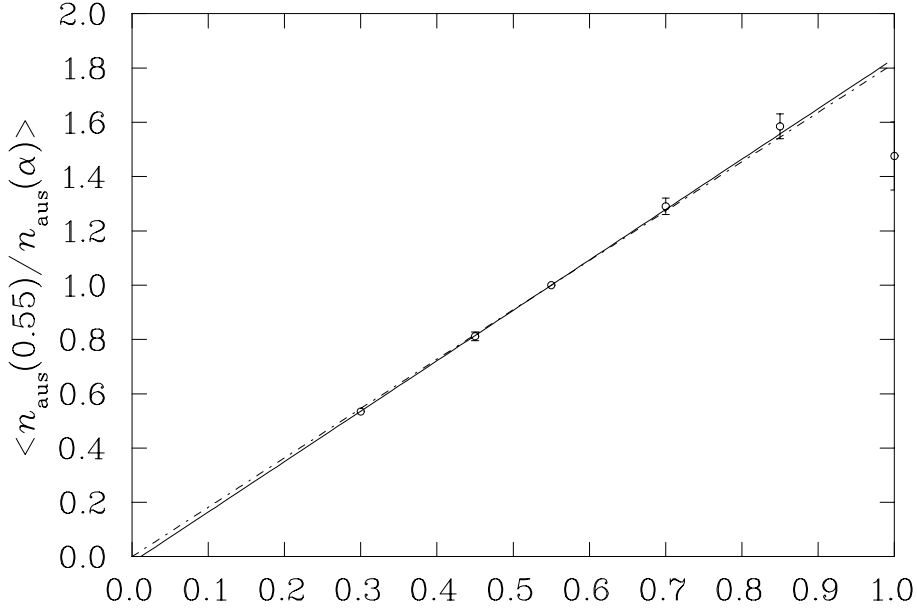,width=12cm}}
\caption{Illustration of the $\alpha$ dependence of $c_1 = \langle
n(\alpha'=0.55)/n(\alpha) \rangle$ for AUS smearing.  Fits to the data
exclude the point at $\alpha=1$.  The solid line fit to the data
indicates $c_1 = 1.857\, \alpha - 0.021$ whereas the dashed line,
constrained to pass through the origin, provides a slope of 1.818.
\label{actaveausfit}}
\end{figure}

Fig.~\ref{actaveapefit} for APE smearing and Fig.~\ref{actaveausfit}
for AUS smearing indicate the relationship between $c_1$ and $\alpha$
is linear with zero intercept to an excellent approximation.  Indeed,
ignoring the point at $\alpha=1$ for AUS smearing, one finds the same
coefficients for the $\alpha$ dependence of APE and AUS smearing.
When the fits are constrained to pass through the origin, one finds a
slope of 1.818 which is the inverse of $\alpha'=0.55$.  Hence we reach
the conclusion that
\begin{equation}
\frac{\napeprm}{\nape} \simeq \frac{\alpha}{\alpha'} \hspace{0.55cm}
{\rm and}
\hspace{0.55cm} 
\frac{\nausprm}{\naus} \simeq \frac{\alpha}{\alpha'} \, .
\label{actrelation}
\end{equation}

This analysis based on the action suggests that a preferred value for
$\alpha$ does not really exist.  In fact it has been recently
suggested that one should anticipate some latitude in the values for
$n(\alpha)$ and $\alpha$ that give rise to effective fat-link actions
\cite{Bernard:1999kc}.  What we have done here is established a
relationship between $n(\alpha)$ and $\alpha$, thus reducing what was
potentially a two dimensional parameter space to a one dimensional
space.  This conclusion will be further supported by the topological
charge analysis below.

\begin{figure}[tp]
\centering{\
\epsfig{angle=90,figure=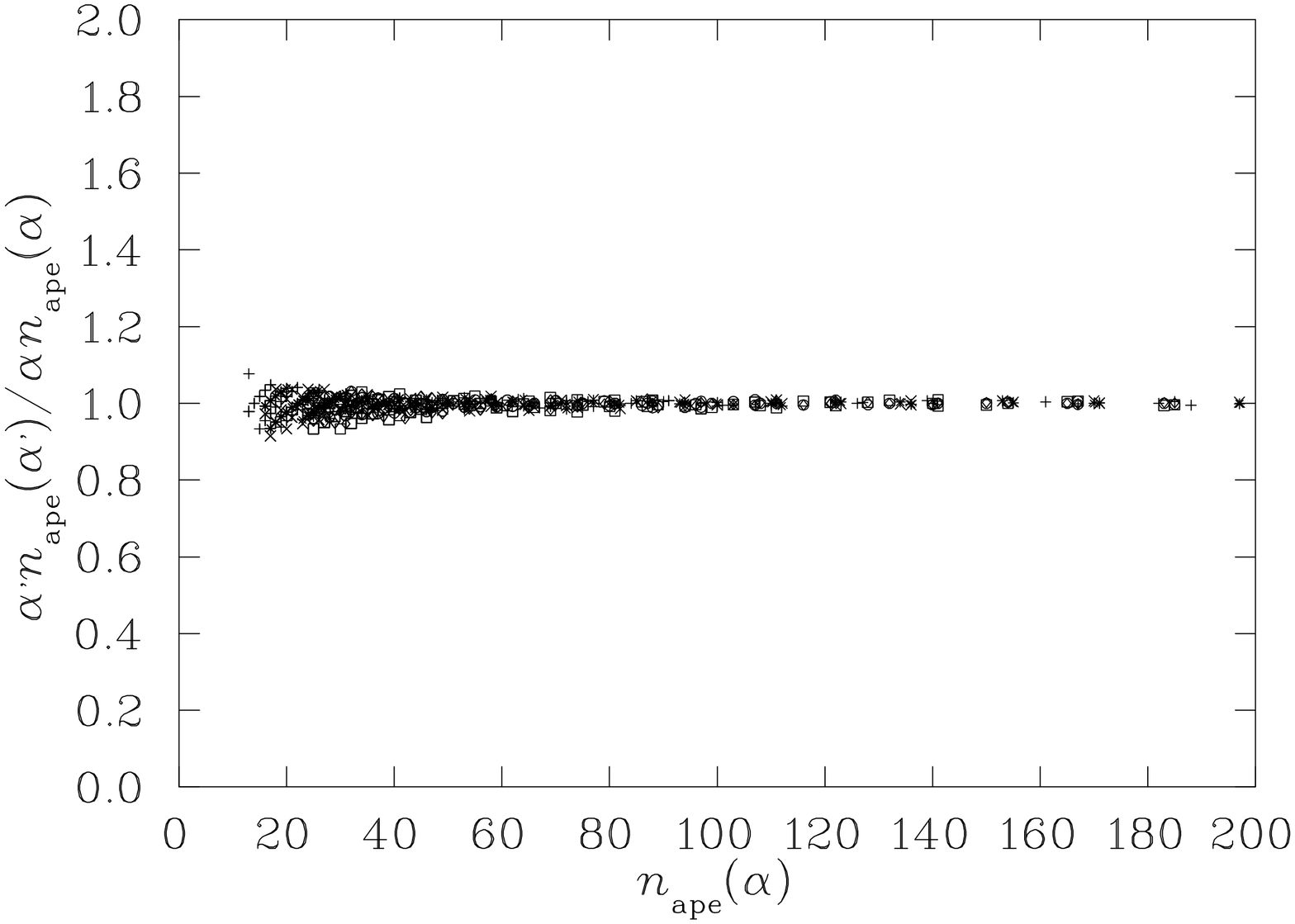,width=12cm}}
\caption{Illustration of the degree to which the relations of
(\protect\ref{actrelation}) are satisfied for the action under APE
smearing.  Here $\alpha < \alpha'$ and $\alpha' = 0.30$, 0.45, 0.55,
and 0.70.  Data are from the $\2$ lattice at $\beta=6.0$.
\label{actaperatio}}
\end{figure}

\begin{figure}[tp]
\centering{\
\epsfig{angle=90,figure=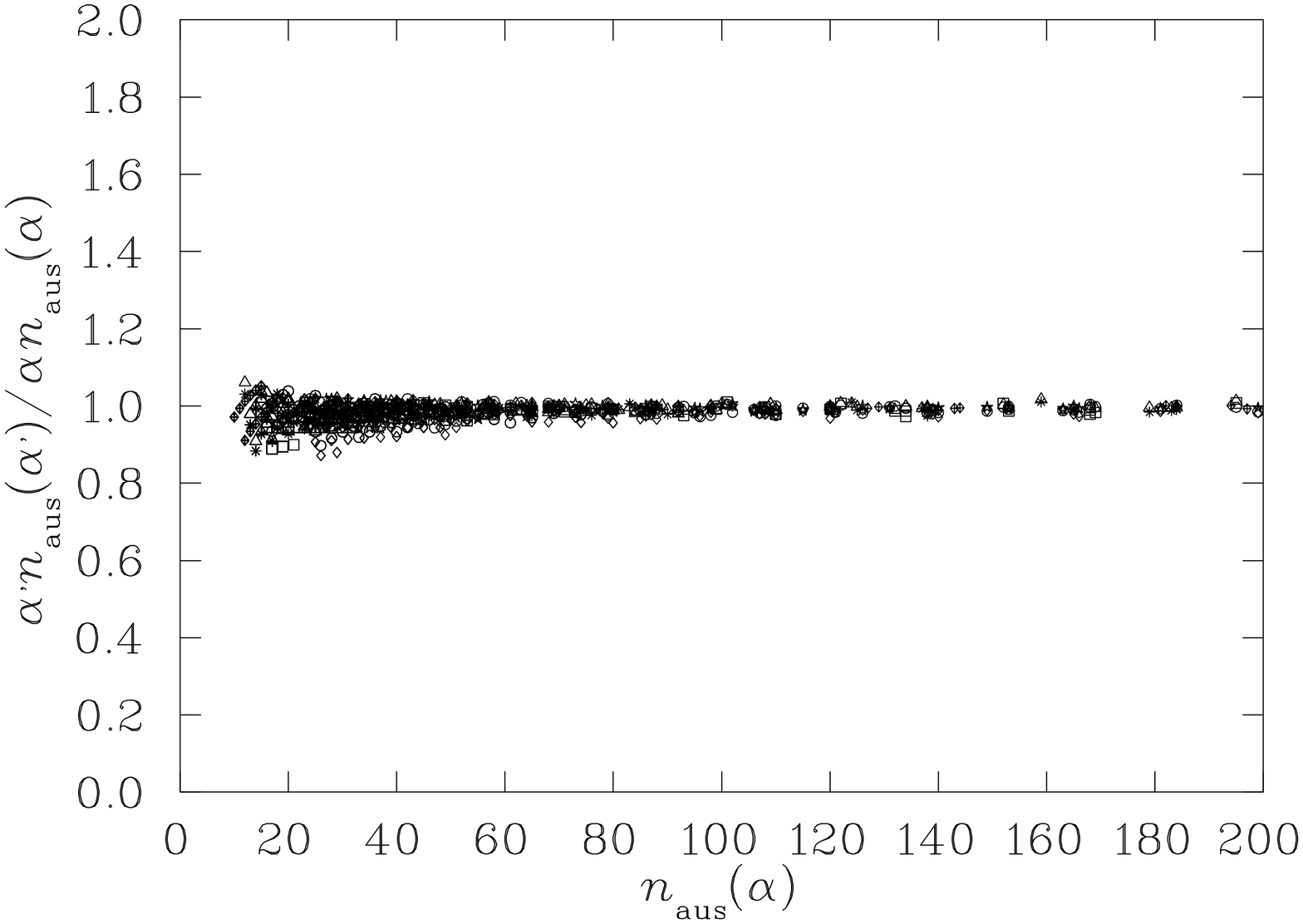,width=12cm}}
\caption{Illustration of the degree to which the relations of
(\protect\ref{actrelation}) are satisfied for the action under AUS
smearing.  Here $\alpha < \alpha'$ and $\alpha' = 0.30$, 0.45, 0.55,
0.70 and 0.85.  Data are from the $\2$ lattice at $\beta=6.0$.
\label{actausrationoal1}}
\end{figure}

\begin{figure}[tp]
\centering{\
\epsfig{angle=90,figure=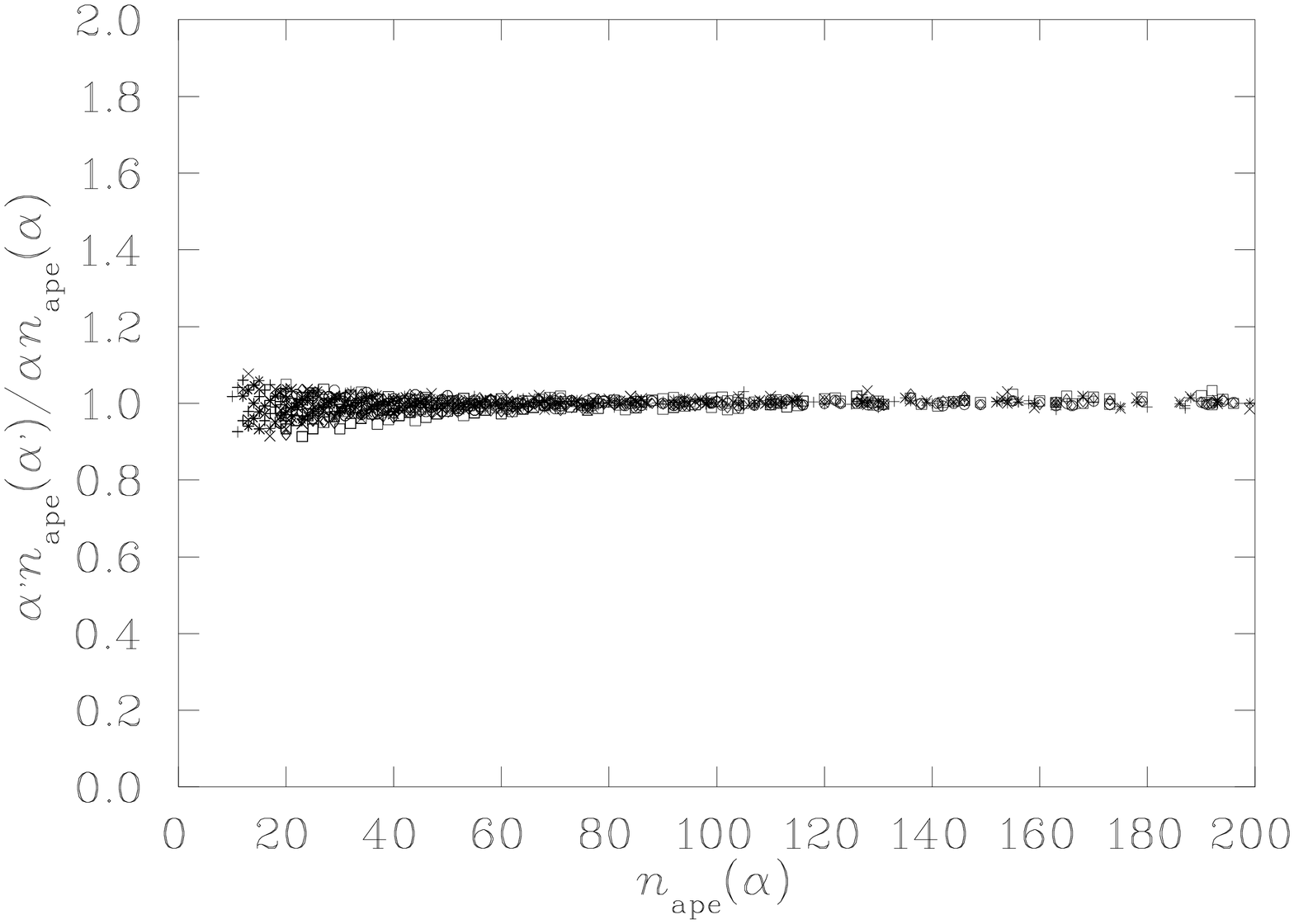,width=12cm}}
\caption{Illustration of the degree to which the relations of
(\protect\ref{actrelation}) are satisfied for the action under APE
smearing.  Here $\alpha < \alpha'$ and $\alpha' = 0.30$, 0.45, 0.55,
and 0.70.  Data are from the $\1$ lattice at $\beta=5.7$.
\label{actaperatioBeta57}}
\end{figure}

\begin{figure}[tp]
\centering{\
\epsfig{angle=90,figure=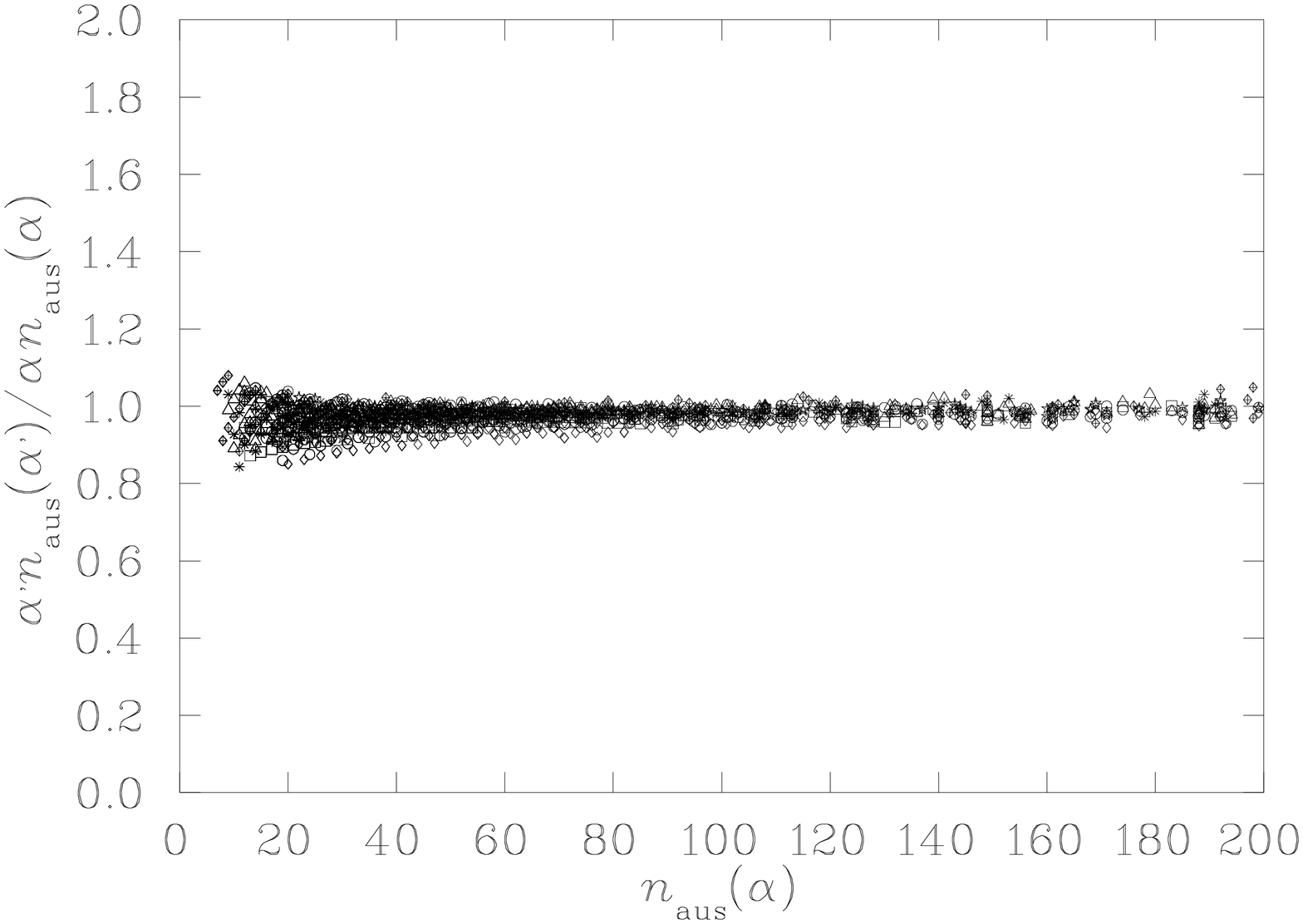,width=12cm}}
\caption{Illustration of the degree to which the relations of
(\protect\ref{actrelation}) are satisfied for the action under AUS
smearing.  Here $\alpha < \alpha'$ and $\alpha' = 0.30$, 0.45, 0.55,
0.70 and 0.85.  Data are from the $\1$ lattice at $\beta=5.7$.
\label{actausrationoal1Beta57}}
\end{figure}

To summarize these finding we plot the ratios
\begin{equation}
\frac{ \alpha' \, \napeprm}{\alpha \, \nape}  \hspace{0.55cm}
{\rm and}
\hspace{0.55cm} 
\frac{\alpha' \, \nausprm}{\alpha \, \naus}  \, ,
\label{ratiosEq1}
\end{equation} 
designed to equal 1 in Figs.~\ref{actaperatio} and
\ref{actausrationoal1} for APE and AUS smearing respectively.
Figs.~\ref{actaperatioBeta57} and \ref{actausrationoal1Beta57} report
the final results of a similar analysis for APE and AUS smearing
respectively at $\beta=5.7$.  Here the $\alpha=1.0$ results are
omitted from the AUS smearing results for clarity.

\subsubsection{Cooling Calibration}

\begin{figure}[tp]
\centering{\
\epsfig{angle=90,figure=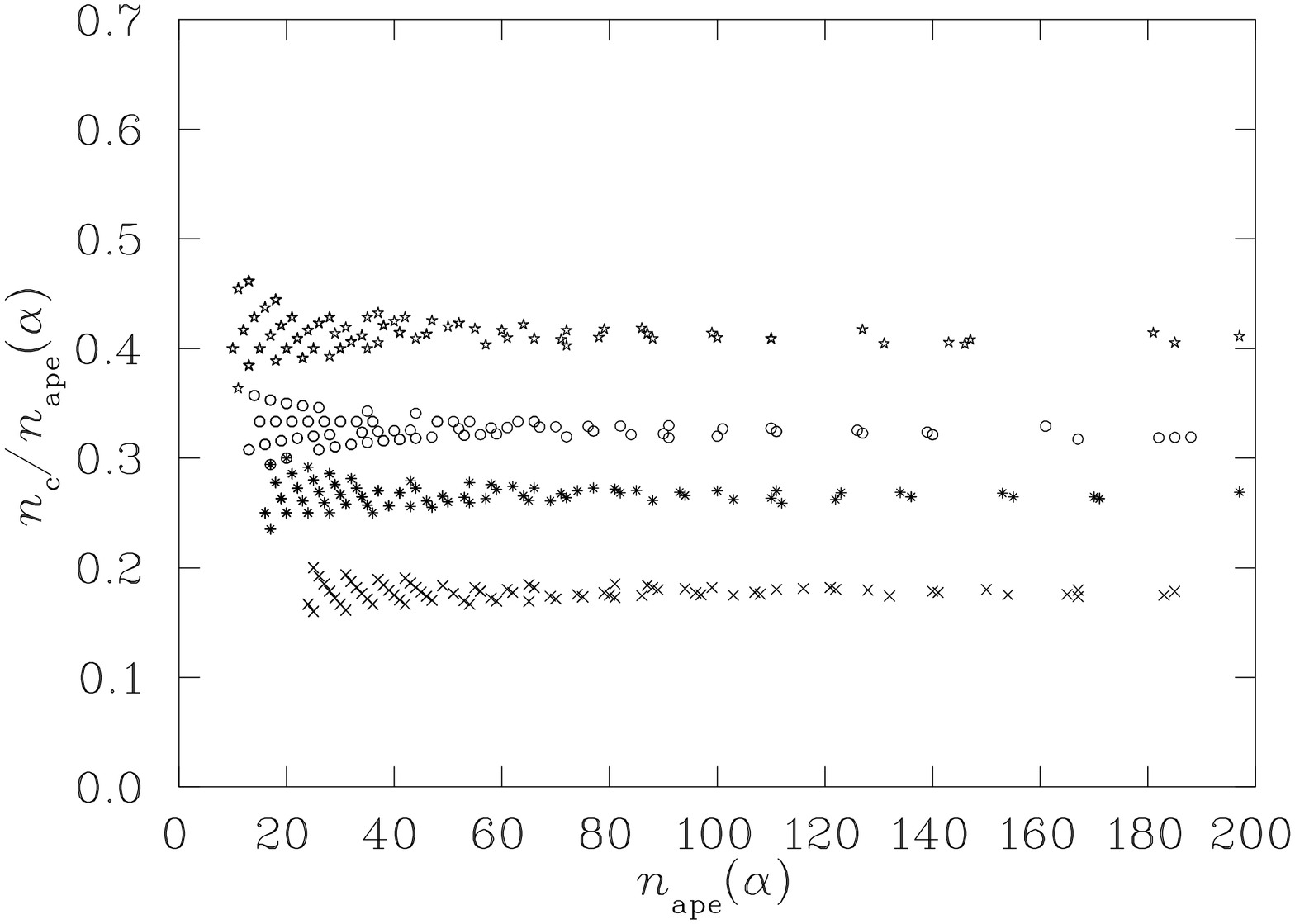,width=12cm}}
\caption{Ratio of cooling to APE smearing sweeps as a function of the
number of APE smearing sweeps.  From top down $\alpha=0.70$ 0.55, 0.45
and 0.30.
\label{actcoolvsapeal}}
\end{figure}

\begin{figure}[tp]
\centering{\
\epsfig{angle=90,figure=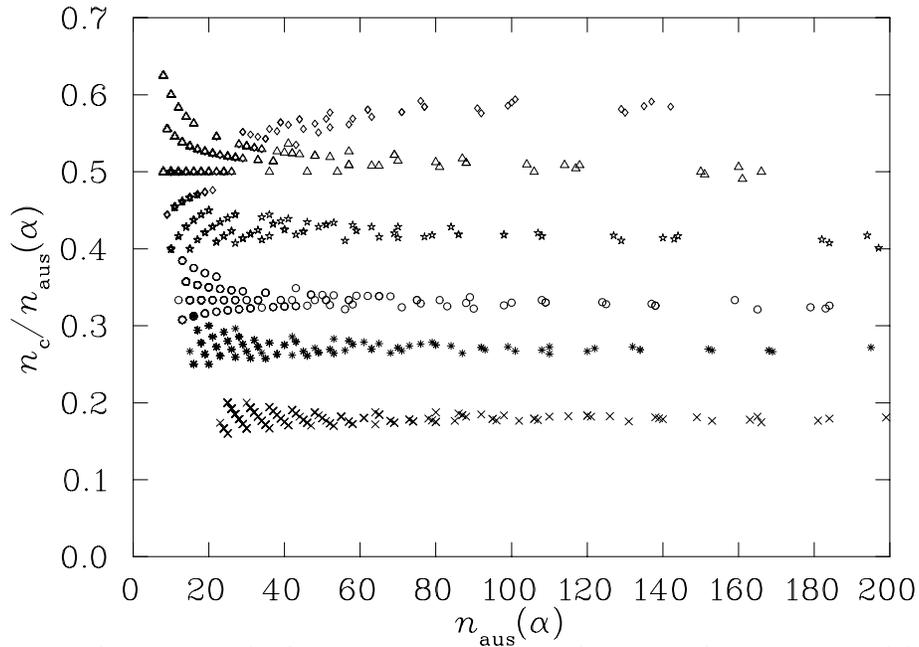,width=12cm}}
\caption{Ratio of cooling to AUS smearing sweeps as a function of the
number of AUS smearing sweeps.  From top down $\alpha=1.00$, 0.85 0.70
0.55, 0.45 and 0.30.
\label{actcoolvsausal}}
\end{figure}

Here we repeat the previous analysis, this time comparing cooling with
APE and AUS smearing.  We make the same linear ansatz for the
relationship and plot $\ncl/\nape$ versus $\nape$ for APE smearing in
Fig.~\ref{actcoolvsapeal}.  Fig.~\ref{actcoolvsausal} reports the
ratio $\ncl/\naus$ versus $\naus$ for AUS smearing.  At small numbers
of smearing sweeps, large integer discretization errors the order of
25\% are present, as $n_c$ is as small as 4.  With this in mind, we
see an independence of the ratio on the amount of cooling/smearing
over a wide range of smearing sweeps for both APE and AUS smearing.
This supports a linear relation between the two algorithms.

Averaging the results provides
\begin{equation}
\frac{\ncl}{\napefv} = 0.330 \hspace*{0.55cm} {\rm and}
\hspace*{0.55cm} \frac{\ncl}{\nausfv} = 0.340 \, ,
\label{ncvssmear}
\end{equation}
or more generally
\begin{equation}
\ncl \simeq 0.600 \, \alpha \, {\nape} 
\hspace*{0.55cm}{\rm and}\hspace*{0.55cm} 
\ncl \simeq 0.618 \, \alpha \, {\naus} \, .
\label{ncforsmear}
\end{equation}
Hence we see that cooling is much more efficient at smoothing than the
smearing algorithms requiring roughly half the number of sweeps for a
given product of $n$ and $\alpha$.

Equating the equations of Eq.~(\ref{ncforsmear}) provides the following relation
between APE and AUS smearing
\begin{equation}
\alpha \, {\nape} \simeq 1.03 \, \alpha' \, {\nausp} \, ,
\label{apeaus}
\end{equation}
summarizing the near equivalence of the two smearing algorithms.  It
is important to recall that the annealing of the AUS smearing
algorithm removes the instability of APE smearing encountered at large
smearing fractions, $\alpha$.  Thus AUS smearing offers a more stable
gauge equivariant smoothing of gauge fields.  It offers faster
smoothing due to its ability to handle larger smearing fractions.
This algorithm may be of use in studying Gribov ambiguities in Landau
gauge fixing \cite{Hetrick:1998}.

A similar analysis of the $\1$ lattice at $\beta = 5.7$ provides
\begin{equation}
\ncl \simeq 0.572 \, \alpha \, {\nape} 
\hspace*{0.55cm}{\rm and}\hspace*{0.55cm} 
\ncl \simeq 0.604 \, \alpha \, {\naus} \, ,
\label{ncforsmear57}
\end{equation}
with
\begin{equation}
\alpha \, {\nape} \simeq 1.06 \, \alpha' \, {\nausp} \, .
\label{apeaus57}
\end{equation}
Here the change in the coefficient relating APE smearing to cooling
appears to be proportional to $\beta$.

\subsection{The Topological Charge Density Analysis.}
\label{numtopo}

\begin{figure}[tp]
\centering{\
\epsfig{angle=90,figure=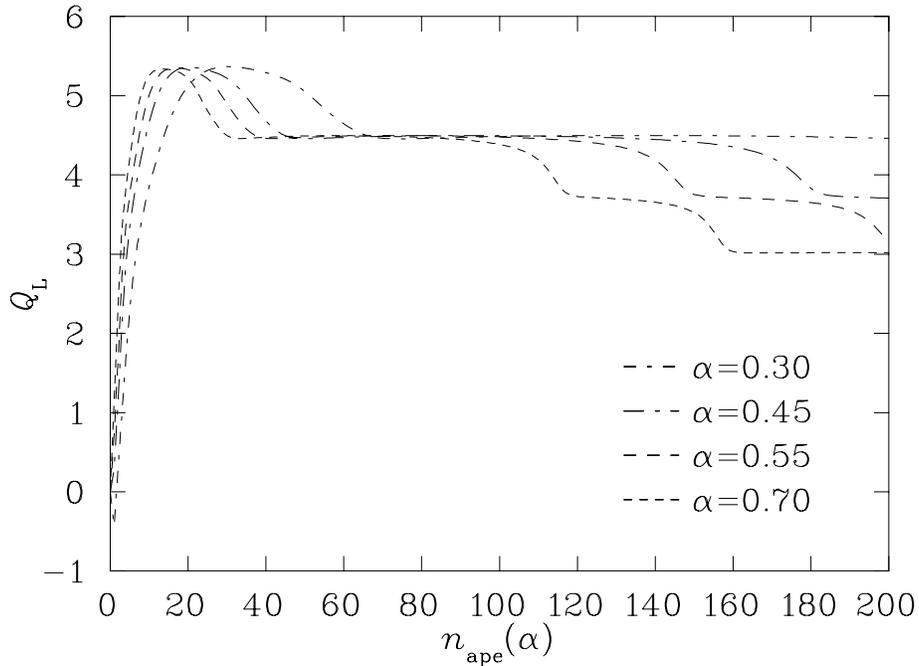,width=12cm} }
\caption{Typical evolution curve of the lattice topological charge
operator as a function of the number of sweeps for APE smearing.  Data
are from the $24^3\times{36}$ lattice at $\beta=6.00$.  Each curve
corresponds to a particular $\al$ value and indicated.
\label{apeevolve}}
\end{figure}

\begin{figure}[tp]
\centering{\
\epsfig{angle=90,figure=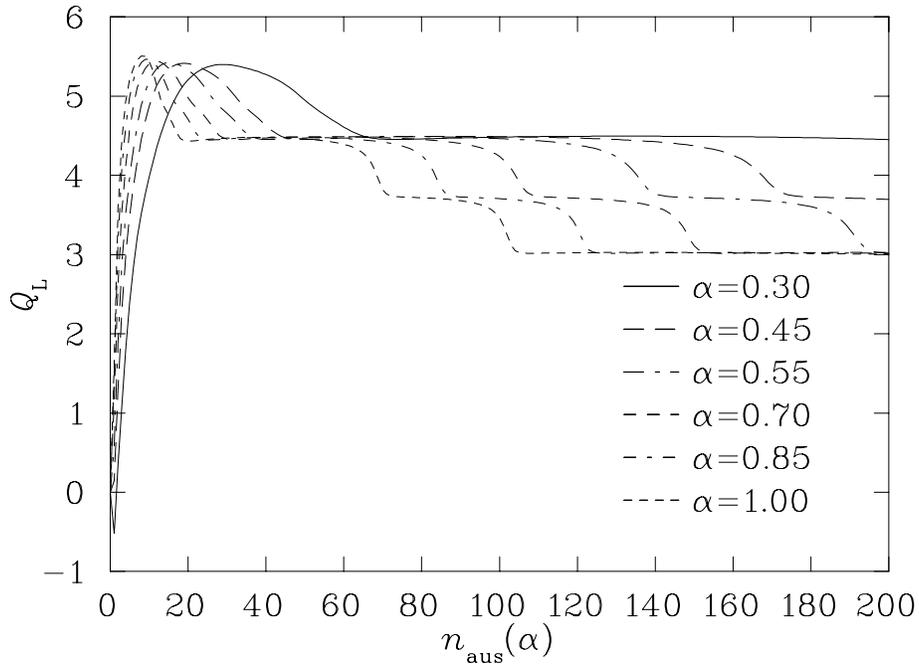,width=12cm} }
\caption{Typical evolution curve of the lattice topological charge
operator under AUS smearing.  Data are from the $24^3\times{36}$
lattice at $\beta=6.00$.  Each curve corresponds to a particular $\al$
value and indicated.
\label{ausevolve}}
\end{figure}

\begin{figure}[tp]
\centering{\
\epsfig{angle=90,figure=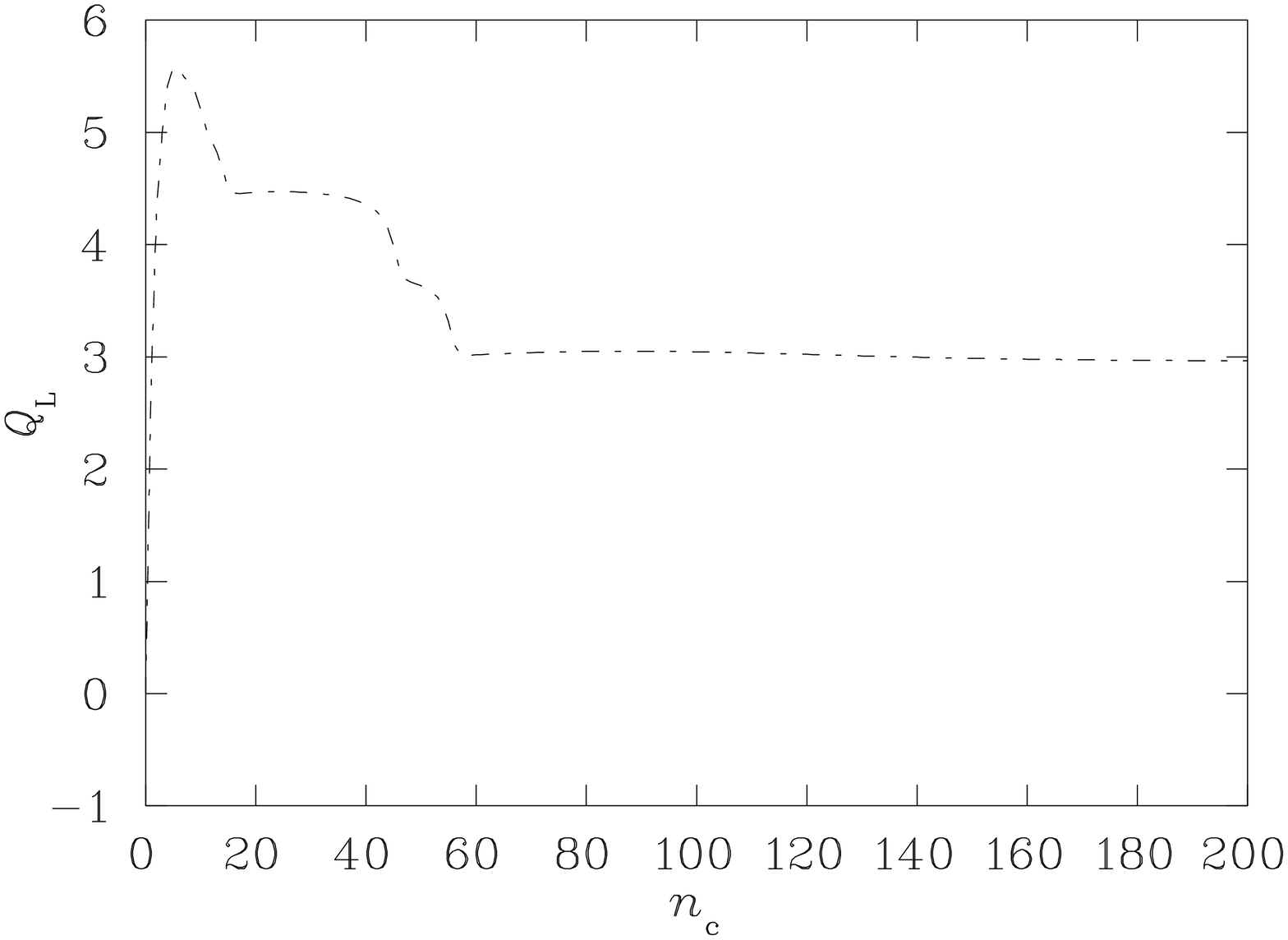,width=12cm} }
\caption{Typical evolution curve of the lattice topological charge
operator as a function of the number of sweeps for cooling with three
diagonal SU(2) subgroups on a $24^3\times{36}$ lattice at
$\beta=6.00$.
\label{coolevolve}}
\end{figure}

Typical evolution curves for the lattice topological charge operator
of Eq.~(\ref{latticecharge}) are shown in Figs.~\ref{apeevolve},
\ref{ausevolve} and \ref{coolevolve} for APE smearing, AUS smearing
and cooling respectively.  These data are obtained from a typical
gauge configuration on our $24^3\times{36}$ lattice at $\beta=6.00$.
The configuration used here is the same representative configuration
illustrated in Figs.~\ref{acapeevolve} and \ref{acausevolve} of the
action analysis.

The main feature of these figures is that the smearing/cooling
algorithms produce a similar trajectory for the topological charge.
For most cases only the rate at which the trajectory evolves changes.
In these cases, one can use these trajectories as another way in which
to calibrate the rates of the algorithms.

After a few sweeps, the topological charge converges to a near integer
value with an error of about 10\%, typical of clover definitions of
$Q$ at $\beta = 6.0$.  The standard Wilson action is known to lose
(anti)instantons during cooling due to ${\cal O}(a^2)$ errors in the
action which act to tunnel through the single instanton action bound.
Given the intimate relationship between cooling and smearing discussed
in Section~\ref{algos} it is not surprising to see similar behavior in
the topological charge trajectories of the smearing algorithms
considered here.  The sharp transitions from one integer to another
indicate the loss of an (anti)instanton.

\begin{figure}[tp]
\centering{\
\epsfig{angle=90,figure=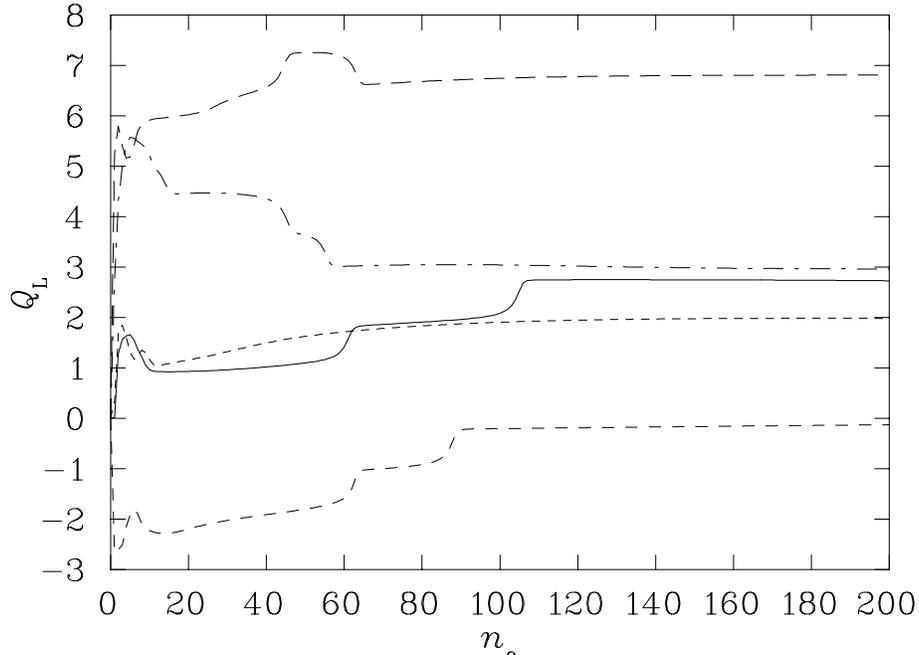,width=12cm}}
\caption{The trajectories of the lattice topological charge density
operator as a function of the number of sweeps for cooling on typical
$24^3\times{36}$, $\beta=6.00$ configurations.
\label{coolevolve5to9}}
\end{figure}

These curves look quite different for other configurations.
Fig.~\ref{coolevolve5to9} displays trajectories for cooling on five
different gauge configurations.  However, the feature of similar
trajectories for the various cooling/smearing algorithms remains at
$\beta = 6.0$.  To calibrate the cooling/smearing algorithms, we
select thresholds at topological charge values where the trajectories
are making sharp transitions from one near integer to another.  Table
\ref{topotab} summarizes the thresholds selected and the number of
sweeps required to pass though the various thresholds.  Note that the
second configuration, $C_{2}$, data entry in Table \ref{topotab}
corresponds to the sampling of the curves illustrated in
Figs.~\ref{apeevolve}, \ref{ausevolve} and \ref{coolevolve}.

\begin{table}[tbp]
\caption{Summary of the number of sweeps required to pass through
various topological charge thresholds.  The selection of the
thresholds is described in the text.  Smearing fractions $\alpha$ are
indicated in the table headings.  Omissions in the table indicate
either the threshold was not met within 200 sweeps or that the
trajectory diverged from the most common trajectory among the
algorithms. }
\begin{tabular}{ccccccccccccc}
Config. &Threshold &Cooling &\multicolumn{4}{c}{APE smearing} &
\multicolumn{6}{c}{AUS smearing} \\
& & & 0.30 & 0.45 & 0.55 & 0.70 & 0.30 & 0.45 & 0.55 & 0.70 & 0.85 & \multicolumn{1}{c}{1.00} \\
\hline
      & 1.00 fall & 2  & 3  & 3  & 2  & -  & 3  & 2  & 2  & -  & -  & \multicolumn{1}{c}{-}   \\
$C_1$ & 1.00 rise & 6  & 7  & 5  & 4  & 3  & 6  & 4  & 3  & 2  & 1  & \multicolumn{1}{c}{1}   \\
      & 1.00 fall & 10 & 41 & 28 & 24 & 20 & 35 & 24 & 19 & 15 & 25 & \multicolumn{1}{c}{17}  \\
      & 1.25 rise & 2  & 9  & 6  & 4  & 3  & 8  & 4  & 3  & 2  & 1  & \multicolumn{1}{c}{1}   \\
      & 1.25 fall & 6  & 25 & 17 & 14 & 12 & 24 & 17 & 13 & 11 & -  & \multicolumn{1}{c}{-}  \\
\hline
      & 5.0 fall & 11 & 52 & 35  & 29  & 23  & 49 & 33  & 26  & 20  & 16  & \multicolumn{1}{c}{13}  \\
      & 4.0      & 45 & -  & 178 & 146 & 114 & -  & 169 & 137 & 110 & 84  & \multicolumn{1}{c}{69}  \\
      & 3.5      & 55 & -  &  -  & 195 & 152 & -  & -   & 189 & 142 & 120 & \multicolumn{1}{c}{101} \\
$C_2$ & 5.0 rise & 3  & 18 & 12  & 10  & 8   & 12 & 11  & 9   & 7   & 5   & \multicolumn{1}{c}{4}   \\
      & 4.0      & 2  & 11 & 7   & 6   & 4   & 10 & 6   & 5   & 3.5 & 3   & \multicolumn{1}{c}{2.5} \\
      & 3.5      & 1  & 8  & 5   & 4   & 3   & 8  & 5   & 4   & 3   & 2   & \multicolumn{1}{c}{2}   \\
      & 3.0      & 1  & 7  & 4   & 3.5 & 3   & 6  & 4   & 3   & 2.5 & 2   & \multicolumn{1}{c}{2}   \\
      & 2.0 rise & 1  & 4  & 3   & 2.5 & 2   & 4  & 3   & 2   & 1.5 & 1   & \multicolumn{1}{c}{1}   \\
\hline
      & 1.0 rise & 2  & 11 & 7   & 6   & 4   & 10 & 6   & 5   & 3  & 2  & \multicolumn{1}{c}{2}  \\
      & 1.0      & 10 & 94 & 62  & 49  & 36  & -  & -   & 26  & 20 & 17 & \multicolumn{1}{c}{16}  \\
$C_3$ & 1.5 rise & 3  & 25 & 17  & 15  & 11  & 21 & 13  & 11  & 8  & 6  & \multicolumn{1}{c}{6}   \\
      & 1.5 fall & 7  & 42 & 25  & 20  & 15  & 39 & 24  & 19  & 15 & 12 & \multicolumn{1}{c}{9}   \\
      & 1.5      & 60 & -  & 168 & 139 & 110 & -  & 162 & 132 & -  & -  & \multicolumn{1}{c}{190} \\
\hline
      & 6.0 & 20 & - & 135 & 110 & 85 & - & 136 & 112 & 88 & - & \multicolumn{1}{c}{-}   \\
$C_4$ & 5.0 & 1  & 6 & 4   & 3   & 2  & - & 4   & 3   & 2  & 2 & \multicolumn{1}{c}{1}  \\
      & 4.0 & 1  & 4 & 3   & 2   & 1  & - & 3   & 3   & 2  & 1 & \multicolumn{1}{c}{1}  \\
\hline
\end{tabular}
\label{topotab}
\end{table}

At $\beta = 5.7$ we did not find analogous trajectories, suggesting
that the coarse lattice spacing and larger errors in the action
prevent one from reproducing a similar smoothed gauge configuration
using different algorithms.  In this case the two-dimensional aspect
of the smearing parameter space remains for studies of the topological
sector.  That this might be the case is hinted at in
Fig.~\ref{actausrationoal1Beta57} in which the ratio of smearing
results for $\beta=5.7$ lattices is not as closely constrained to one
as for the $\beta=6.0$ results in Figs.~\ref{actausrationoal1}.

\begin{figure}[tp]
\centering{\
\epsfig{angle=90,figure=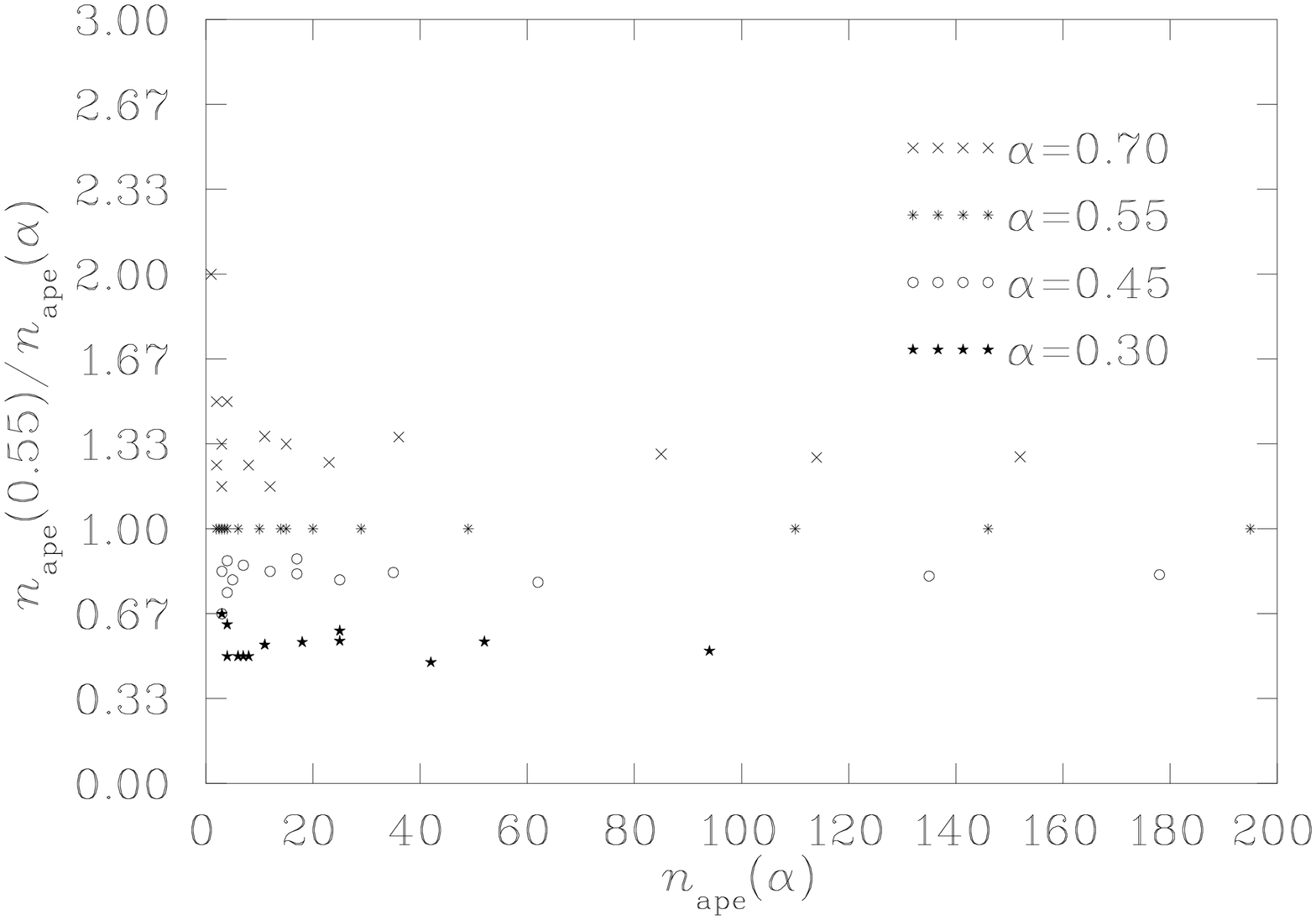,width=12cm}}
\caption{The ratio $\napefv/\nape$ versus $\nape$ for $Q_{L}$
from the data of Table \protect\ref{topotab} extracted from the $\2$
configurations at $\beta=6.00$.  From the top down, the horizontal sets of
points correspond to $\al=0.70$, 0.55, 0.45 and 0.30. 
\label{topoape055vsapeal}}
\end{figure}

\begin{figure}[tp]
\centering{\
\epsfig{angle=90,figure=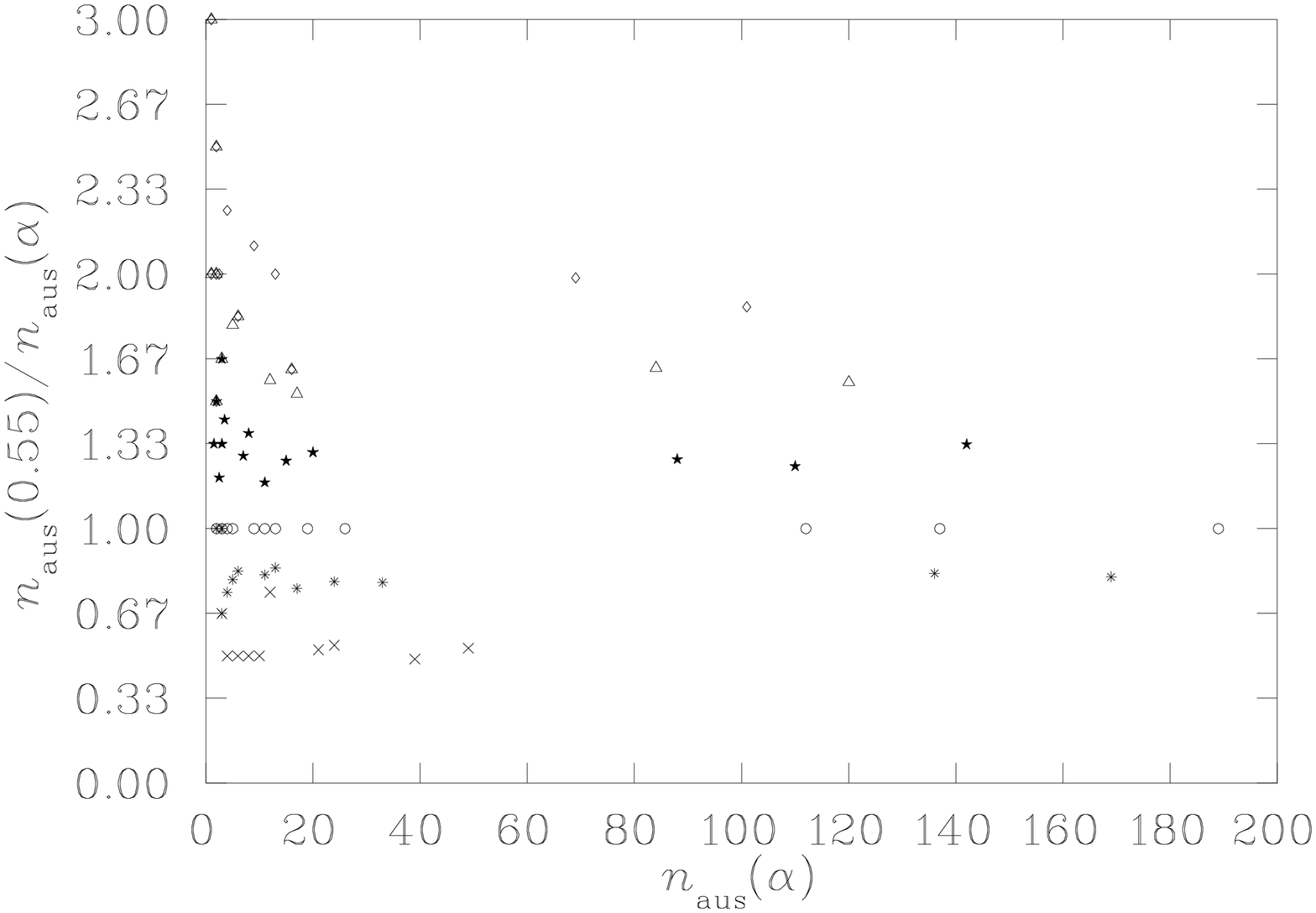,width=12cm}}
\caption{The ratio $\nausfv/\naus$ versus $\naus$ for $Q_{L}$
from the data of Table \protect\ref{topotab} extracted from the $\2$
configurations at $\beta=6.00$.  From the bottom up, the symbols
correspond to $\al=0.30$, 0.45, 0.55, 0.70, 0.85 and 1.00. 
\label{topoaus055vsausal}}
\end{figure}

As in the action analysis, we report the ratio of $\al=0.55$ results
to other $\alpha$ value results within APE and AUS smearing.
Figs.~\ref{topoape055vsapeal} and \ref{topoaus055vsausal} summarize
the results for APE and AUS smearing respectively.  Again we see an
enhanced spread of points at small numbers of smearing sweeps due to
integer discretization errors.

Data for large numbers of sweeps at small and large $\alpha$ values
are absent in Fig.~\ref{topoaus055vsausal}.  The absence of points for
small $\alpha$ values is simply due to the thresholds not being
crossed within 200 smearing sweeps.  The absence of points for large
$\alpha$ values reflects the divergence of the topological charge
evolution from the most common trajectory among the algorithms.  This
divergence is also apparent in Fig.~\ref{actaus055vsal} where the
$\alpha = 1.0$ AUS smearing results fail to satisfy the linear ansatz.

While the data from the topological charge evolution is much more
sparse, one can see reasonable horizontal bands forming supporting a
dominant linear relationship between various $\alpha$ values.
Averages of the bands are reported in Table~\ref{avetab} along with
previous results from the action analysis.  With the exception of the
$\alpha=1.0$ AUS smearing results, the agreement is remarkable,
leading to the same conclusions of the action analysis summarized in
Eq.~(\ref{actrelation}).

\begin{table}[b]
\caption{The average of the ratio $<\napefv/\nape>$ or
$<\nausfv/\naus>$ for the action analysis, $(S)$, and the topological
charge analysis, $(Q)$ from the $\2$ lattice.}
\begin{tabular}{ccccccccccc}
    & \multicolumn{4}{c}{APE smearing} & \multicolumn{6}{c}{AUS smearing} \\
\hline
$\al$   & 0.30 & 0.45 & 0.55 & 0.70 & 0.30 & 0.45 & 0.55 & 0.70 & 0.85 & 1.00 \\
$S$ & 0.541(1) & 0.816(1) & 1.0 & 1.278(1) & 0.534(1) &
0.812(1) & 1.0 & 1.290(1) & 1.584(2) & 1.475(7) \\ 
$Q$ & 0.54(1) & 0.83(1) & 1.0 & 1.28(2) & 0.54(3) & 0.812(8) &
1.0 & 1.28(1) & 1.65(4) & 1.90(6) \\ 
\end{tabular}
\label{avetab}
\end{table}

\begin{figure}[tp]
\centering{\
\epsfig{angle=90,figure=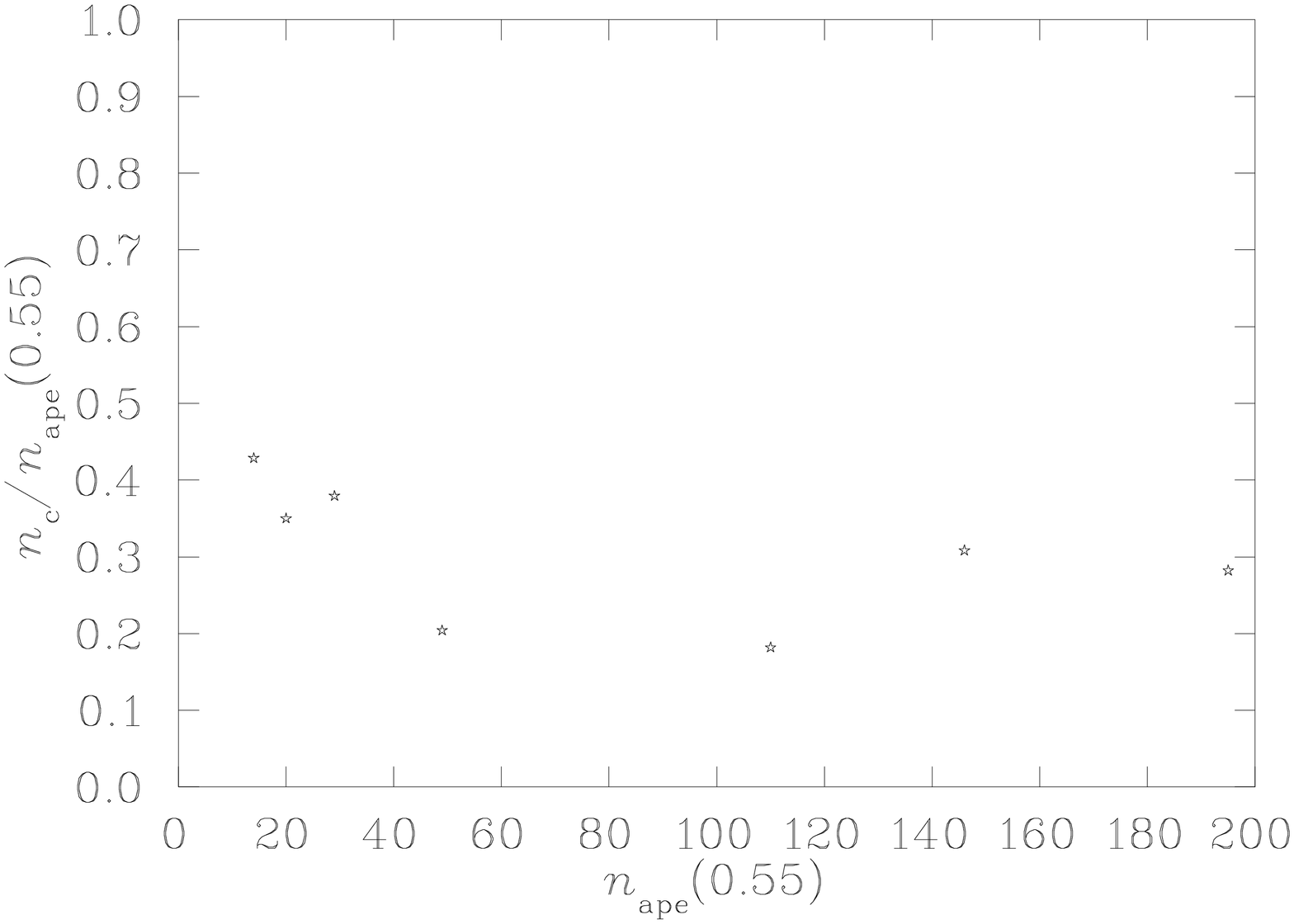,width=12cm}}
\caption{Calibration of cooling and APE smearing via the ratio
$\ncl/\napefv$ versus $\napefv$.
\label{coolvsapefit}}
\end{figure}

\begin{figure}[tp]
\centering{\
\epsfig{angle=90,figure=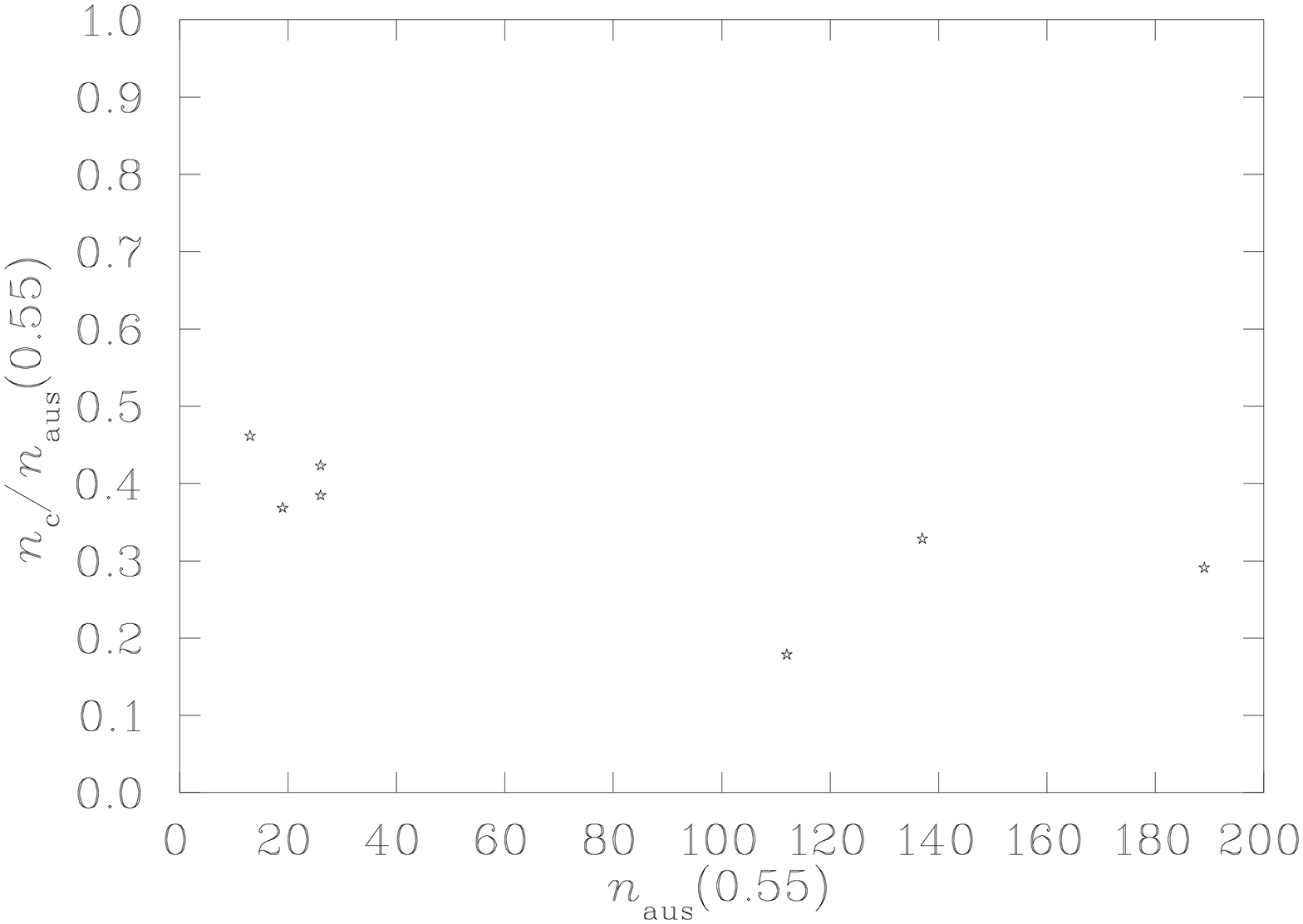,width=12cm}}
\caption{Calibration of cooling and AUS smearing via the ratio
$\ncl/\nausfv$ versus $\nausfv$.
\label{coolvsausfit}}
\end{figure}

In calibrating cooling via the topological charge defects, we report
the ratio $\ncl/\napefv$ as a function of $\nape$ and 
$\ncl/\nausfv$ as a function of $\naus$ for APE and AUS smearing
respectively.  The results are shown in Figs.~\ref{coolvsapefit} and
\ref{coolvsausfit}.

The data is too poor to determine anything beyond a linear relation
between $\ncl$ and $\nape$ or $\naus$.  Averaging the results provides 
\begin{equation}
\frac{\ncl}{\napefv} = 0.30(3) \hspace*{0.55cm} {\rm and}
\hspace*{0.55cm} \frac{\ncl}{\nausfv} = 0.35(3) \, ,
\label{ncvssmeartop}
\end{equation}
in agreement with the earlier action based results of
Eq.~(\ref{ncvssmear}).

\begin{figure}[tp]
\centering{\
\epsfig{figure=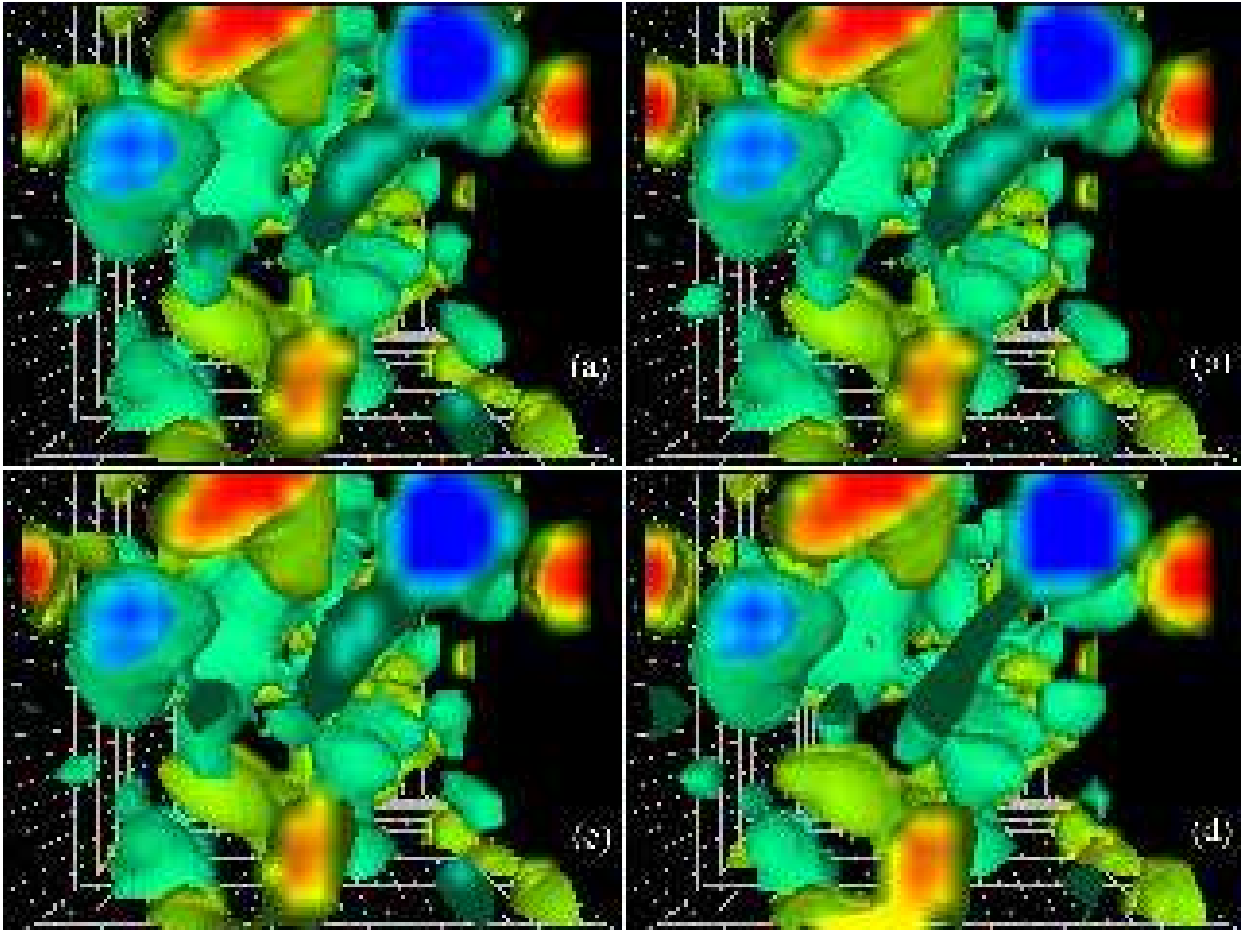,width=\hsize}}
\vspace{12pt}
\caption{Topological charge density of a $\2$ lattice for fixed $x$
coordinate.  Positive (negative) windings are colored red to yellow
(blue to green).  Fig.~(a) illustrates the topological charge density
after 21 APE smearing steps at $\alpha = 0.7$.  Fig.~(b) illustrates
the topological charge density after 49 APE smearing steps at $\alpha
= 0.3$.  Relation (\protect\ref{actrelation}) suggests these
configurations should be similar and we see they are to a remarkable
level of detail.  Fig.~(c) illustrates the topological charge density
after 20 AUS smearing steps at $\alpha = 0.7$.  Relation
(\protect\ref{apeaus}) suggests these configurations should be similar
and again the detail of the agreement is excellent.  Finally (d)
illustrates the topological charge density after 9 cooling sweeps,
motivated by relation (\protect\ref{ncforsmear}).
While the level of agreement is not as precise, the qualitative
features of the smoothed configurations are compatible.
\label{SmearingComp}}
\end{figure}

As a final examination of the relations we have established among APE
smearing, AUS smearing and cooling algorithms, we illustrate the
topological charge density in Fig.~\ref{SmearingComp}.  Large positive
(negative) winding densities are shaded red (blue) and symmetric
isosurfaces aid in rendering the shapes of the densities.  In fixing
the $x$ coordinate to a constant, a three-dimensional slice of a
four-dimensional $\2$ gauge field configuration is displayed.
Fig.~\ref{SmearingComp}(a) illustrates the topological charge density
after 21 APE smearing steps at $\alpha = 0.7$.  The other three
quadrants display APE smearing, AUS smearing and cooling designed to
reproduce Fig.~\ref{SmearingComp}(a) according to the relations of
Eqs.~(\protect\ref{actrelation}), (\protect\ref{ncforsmear}) and
(\protect\ref{apeaus}).  The level of detail in the agreement is
remarkable.

\section{Conclusion}
\label{summary}

The APE smearing algorithm is now widely used in a variety of ways in
lattice simulations.  It is used to smear the spatial gauge-field
links in studies of glueballs, hybrid mesons, the static quark
potential, etc.  It is used in constructing fat-link actions and in
constructing improved operators with smooth transitions to the
continuum limit.  We have shown that to a good approximation the
two-dimensionful parameter space of the number of smearing sweeps
$\nape$ and the smearing fraction $\alpha$ may be reduced to a single
dimension via the constraint
\begin{equation}
\frac{\napeprm}{\nape} = \frac{\alpha}{\alpha'} \, .
\label{APEsummary}
\end{equation} 
satisfied for $\alpha$ and $\alpha'$ in the range 0.3 to 0.7.  This
result is in agreement with fat-link perturbation theory expectations,
and survives for up to 200 sweeps over the lattice.  We expect this
relation to hold over the entire APE smearing range $0 < \alpha <
3/4$.  We find the same relation for AUS smearing provided $\alpha \le
0.85$.  For AUS smearing, annealing of the links is included as one
cycles through the Lorentz directions of the link variables in the
smearing process.

The smoothing of gauge field configurations is often a necessary step
in extracting observables in which the renormalization constants
differ significantly from one, as is the case for the topological
charge operator.  It is also often used to gain insights into the
nonperturbative features of the field theory which give rise to the
observed phenomena.  We have determined that cooling, APE smearing and
AUS smearing produce qualitatively similar smoothed gauge field
configurations at $\beta=6.0$ provided one calibrates the algorithms
as follows:
\begin{equation}
\ncl \simeq 0.600 \, \alpha \, {\nape} \, ,
\hspace*{0.55cm} 
\ncl \simeq 0.618 \, \alpha \, {\naus} \,  
\hspace*{0.55cm} {\rm and} \hspace*{0.55cm} 
\alpha \, {\nape} \simeq 1.03 \, \alpha' \, {\nausp} \, .
\label{summaryeqs}
\end{equation}

The topological charge analysis serves to confirm the action analysis
results at $\beta=6.0$ and further support these relations.  However,
it also reveals that at $\beta=5.7$ different trajectories are taken.
At $\beta=5.7$ different algorithms produce smoothed gauge field
configurations with similar action, but different topological
properties.  

As most modern simulations are performed at $\beta \ge 6.0$, the
relations of Eqs.~(\ref{APEsummary}) and (\ref{summaryeqs}) will be
most effective in reducing the exploration of the parameter space.  It
is now possible to arrive at optimal smearing by fixing the number of
smearing sweeps and varying the smearing fraction, or vice versa.
Finally, it is now possible to directly compare the physics of smeared
and cooled gauge field configurations in a quantitative sense.

We have used the action and topological charge trajectories to
calibrate the various smoothing algorithms.  While the topological
charge may be related to the topological susceptibility, it may be of
future interest to introduce other physical quantities such as the
string tension as a measure of smoothing \cite{garcia99}.

It is well known that interpreting the physics of cooled
configurations based on the standard Wilson action is somewhat of an
art.  The difficulty is that the ${\cal O}(a^2)$ errors in the action
spoil an instanton under cooling by reducing the action below the
one-instanton bound.  A consequence of this is a lack of universality.
For example, the action varies smoothly to zero while the topological
sector jumps as (anti)instantons are destroyed.  One may then attempt
to define the concept of an optimal amount of smoothing, and perhaps
also extrapolate back to zero smoothing steps.  Unfortunately, all of
these difficulties are apparent in the APE smearing analysis results
as well.

Because of the intimate connection between smearing and cooling, as
emphasized in the discussion of Section \ref{aussmear}, it is natural to
consider improved smearing, where the Hermitian conjugate of the
improved staple is used in the smearing process.  It will be
interesting to see if the benefits of improved cooling carry over into
improved smearing and this work is currently in progress. 

\section*{Acknowledgment}

Thanks are extended to Tom Degrand for helpful comments on fat-link
perturbation theory.  Thanks also to Francis Vaughan of the South
Australian Centre for Parallel Computing and the Distributed
High-Performance Computing Group for support in the development of
parallel algorithms implemented in Connection Machine Fortran (CMF)
and for generous allocations of time on the University of Adelaide's
CM-5.  Support for this research from the Australian Research Council
is gratefully acknowledged.  AGW also acknowledges support form the
Department of Energy Contract No.~DE-FG05-86ER40273 and by the Florida
State University Supercomputer Computations Research Institute which
is partially funded by the Department of Energy through contract
No.~DE-FC05-85ER25000.

\end{document}